\documentclass[12pt]{iopart}
\usepackage{color}
\usepackage{graphicx}
\usepackage{epsf}
\usepackage{graphicx,epsfig}
\usepackage{tabularx}
\usepackage{epstopdf}
\usepackage{axodraw4j}
\usepackage{pstricks}
\usepackage{amsssymb}
\usepackage{simplewick}
\newcommand{\Rmnum}[1]{\expandafter\@slowromancap\romannumeral #1@}

\def\cs2{c_{s}^{2}}

 \def\be   {\begin{equation}}   \def\ee   {\end{equation}}
 \def\ba   {\begin{array}}      \def\ea   {\end{array}}
 \def\bea  {\begin{eqnarray}}   \def\eea  {\end{eqnarray}}
 \def\bean {\begin{eqnarray*}}  \def\eean {\end{eqnarray*}}

\begin{document}

\title{Large non-Gaussianities in the Effective  Field Theory  Approach to Single-Field Inflation:\\ the Trispectrum}

\author{Nicola Bartolo$^{1,2}$, Matteo Fasiello$^{3}$, Sabino Matarrese$^{1,2}$ and 
Antonio Riotto$^{2,4}$}
\vspace{0.4cm}
\address{$^1$ Dipartimento di Fisica ``G. Galilei'', Universit\`{a} degli Studi di 
Padova,  via Marzolo 8, I-35131 Padova, Italy} 
\address{$^2$ INFN, Sezione di Padova, via Marzolo 8, I-35131 Padova, Italy}
\address{$^3$ Dipartimento di Fisica ``G. Occhialini'', Universit\`{a} degli Studi di Milano Bicocca and INFN, Sezione di Milano Bicocca, Piazza della Scienza 3, I-20126 Milano, Italy}
\address{$^4$ CERN, Theory Division, CH-1211 Geneva 23, Switzerland\\
\vskip 0.5cm
DFPD-2010-A-13\,\, CERN-PH-TH/2010-146}
\vskip 0.5cm
\eads{\mailto{nicola.bartolo@pd.infn.it}, \mailto{matteo.fasiello@mib.infn.it}, 
\mailto{sabino.matarrese@pd.infn.it} and \mailto{riotto@mail.cern.ch}}

\begin{abstract}
We perform the analysis of the trispectrum of curvature perturbations generated by the interactions characterizing a general theory of single-field inflation obtained by effective field theory methods. We find that curvature-generated interaction terms, which can in general give an important contribution to the amplitude of the four-point function, show some new distinctive features in the form of their trispectrum shape-function. These interesting interactions are invariant under some recently proposed symmetries of the general theory and, as shown explicitly, do allow for a large value of the trispectrum.
\end{abstract}
\newpage

\tableofcontents

\section{Introduction}
Inflation \cite{guth} is one of the central pillars of modern cosmology. Not only it provides a natural solution to the flatness, horizon and monopole problems of standard Big-Bang cosmology, but can also explain the production of density perturbations in the early Universe which then lead to LSS \cite{lss1,lss2,lss3,lss4,lss5} in the distribution of galaxies and temperature anisotropies in the CMB \cite{smoot92,bennett96,gorski96,wmap3,wmap5,kom}. \\
Besides the simplest single-field slow roll inflationary model, many other inflationary mechanisms have been proposed since inflation was first introduced, and all are compatible with the CMB and LSS observations. In order to probe deeper into the dynamics of inflation and to remove the degeneracies generated by the many models so far proposed, one might study observable quantities which are sensitive to deviations from Gaussianity \cite{reviewk}: starting from the three-point function \cite{bisp,bispectrum}, one then considers the trispectrum \cite{trispectrum} and in general higher-order correlators as well as loop effects in the power spectrum \cite{loop}. Such non-Gaussian features will depend on the various interactions characterizing any given inflationary model in the form of self-interactions of the inflaton, its coupling with gravity and interactions with other fields in the case of multi-field inflation (see, e.g., \cite{koyamarev,ourrev,chenrev} for comprehensive and updated reviews). These investigations are spurred by the fact that the continued analysis of WMAP data~\cite{kom} and the recent launch of the Planck satellite~\cite{Pl,Mand} provide the exciting opportunity to actually test the predictions of this zoo of models at the level of bispectrum and trispectrum of curvature perturbations. 
A very useful tool in analyzing the possible signatures of the different inflationary models is given by the effective field theory approach to inflation recently introduced in \cite{luty} and further expanded in \cite{eft08,3pt,w-e,ssz05}. There are various advantages in employing this formalism. Indeed, it provides a unifying perspective on inflation in that it automatically accounts for many known inflationary mechanisms. To each set of interactions for a given inflationary model there corresponds in the effective Lagrangian a linear combination of operators obtained by turning on and off some specific coefficients that regulate the weight of the operators (we introduce these coeffcients later on and refer to them as ${\bf M}_n$'s). The unifying power of the effective field theory approach is quite manifest in that, in principle, it allows these coefficients considerable more freedom than what they are granted in any specific inflationary model. In fact, by being for the most part free parameters (a couple of these coefficients are to obey some inequalities if one wants, as we do, the generalized speed of sound to be smaller than unity), the ${\bf M}_n$'s allow for the description of known interactions with relative weights which would otherwise be fixed, so by employing effective field theory one enlarges the region of the parameters space than can be spanned. Besides that, in the effective Lagrangian some of the ${\bf M}_n$'s multiply curvature-generated operators (see below Eq.~(\ref{unper})) that are sometimes neglected but should in principle be studied as, in fact, their contribution can be relevant \cite{b} and increase the dimension of the parameters space of the theory.\\
\noindent In this paper we use effective field theory techniques to study the trispectrum of a very general theory for single field inflation. In particular, we concentrate on the contributions of many novel curvature-generated terms as well as interactions that characterize Ghost inflation \cite{ghost}. As an ordering principle among the numerous interactions such a general approach comprises, we employ two additional symmetries of the action recently introduced in \cite{muko,4pt} (see also \cite{huang}) and examine the cases when one or both of these requirements are imposed on the theory. Following \cite{chen-tris}, we analyze the shapes of these terms in four different configurations so as to identify distinctive effects in the trispectrum shape-function from the various interactions. We first analyze the contribution to the scalar exchange diagram due to a curvature-related term which generates an interesting flat shape for the bispectrum \cite{b}. This term produces a shape function which is, in some configurations, different from all the shapes due to leading interactions in \textit{general single-field inflation}  models \cite{chen-tris}.\footnote{In these models the inflaton Lagrangian is an arbitrary function of the inflaton and its \textit{first} derivative. The theories we will consider here further generalize these models. } We calculate and plot the contributions to the contact interaction diagram by several interactions: we rediscover the shape-function due to the leading fourth-order interaction in ghost inflation first obtained in \cite{muko} and plot it in new configurations; an analysis of other interesting term is also performed allowing us to identify novel distinctive features of curvature-related interactions.\\
The paper is organized as follows. In section 2 we build on \cite{eft08} to introduce the very general effective theory we employ in all subsequent calculations. The reader who is familiar with this procedure might want to skip this part and start directly from Eq.~(\ref{h4}). In section 3 we characterize the various interaction terms of the theory according to their behaviour under the action of two specific symmetries. 
Afterwards, in section 4, we proceed by briefly outlining the tools of the IN-IN formalism successively employed in the trispectrum calculations. A separate analysis of the scalar exchange and contact interaction diagrams contributions to the trispectrum is performed. Section 5 contains a summary of the findings and comments on further work. In \textit{Appendix A} we report the details of the scalar exchange diagram calculations. In \textit{Appendix B} we show the reliability of our simplifying assumption on the classical solution to the equations of motion for the general theory considered in this paper.\\
For the sake of clarity, we stress here that whenever we refer to \textit{general single-field inflation} models, often when elaborating on the results of \cite{chen-tris}, we are dealing with theories which account for a great fraction of known inflationary models (DBI, K-inflation etc.), but still miss an important subset including Ghost inflation and curvature-generated interactions in general. On the other hand, the effective approach of \cite{eft08} we employ here also covers the latter. 
\section{The Hamiltonian up to fourth order}
We will follow the  effective theory approach first introduced in  Ref.  \cite{eft08} in order to write down the complete theory of single-field models of inflation   up to fourth order in perturbations, subjected to the sole requirement of an approximate shift symmetry, $\pi\rightarrow \pi + \textit{const}$, on the scalar degree of freedom  $\pi$. Let us give a brief account on how to obtain the main formulas. We start with the scalar field $\phi$ responsible for inflation, which is split as an unperturbed part plus the fluctuation:
\be
\phi(\vec x, t)=\phi_0 (t)+ \delta \phi(\vec x, t).
\ee 
For reasons that will soon become clear, one chooses here to work in the the comoving (or unitary) gauge for which $\delta \phi =0$  \cite{luty}. As a result,  the 
 action will no longer be invariant under full spacetime diffeomorphisms (diffs) but only under the spatial reparametrizations.
 This is the starting point to write the most general unitary gauge space diffs invariant Lagrangian at the desired order in perturbation theory \cite{eft08}:

\be
S = \int d^4 x \sqrt{-g} \; F(R_{\mu\nu\rho\sigma}, g^{00}, K_{\mu\nu},\nabla_\mu,t),\label{unper}
\ee
where $K_{\mu\nu}$ is the extrinsic curvature and the indices on the metric entries $g^{00}$ are free indices.  Taking into account the fluctuations around a FRW background, one obtains the following action
\begin{eqnarray}
S &=& \int d^4x \sqrt{-g} \; \Big[\frac12 M_{\rm Pl}^2 R + M_{\rm Pl}^2  \dot H  g^{00} - M_{\rm Pl}^2 \Big(3 H^2 +\dot H\Big) + \nonumber \\  && 
\sum_{n\geq 2} F^{(n)}(g^{00}+1,\delta K_{\mu\nu}, \delta R_{\mu\nu\rho\sigma};\nabla_\mu;t)\Big],
\label{genspacei}
\end{eqnarray}
where the fluctuations contained in $F^{(n)}$ are at least second order.
The next step is restoring full spacetime diffs invariance. To see how it works, we take from \cite{eft08} the following sample terms in the action:
\be
\int d^4x\;  \sqrt{-g} \left[A(t)+B(t)g^{00}(x)\right] \ \label{ex1}.
\ee
Consider the time reparametrization: $t\rightarrow t+\xi_{0}(\vec x,t); \quad \vec x \rightarrow \vec x $;  under its action (and after a simple variable redefinition) Eq.~(\ref{ex1}) reads:
\be
\fl \int d^4x\;  \sqrt{- g(x)} \left[A( t-\xi_0(x))+B(t-\xi_0(x)) \frac{\partial (t-\xi_0(x))}{\partial x^\mu}\frac{\partial (t-\xi_0(x))}{\partial x^\nu} g^{\mu\nu}(x)\right].\label{ex2}
\ee
At this stage the procedure we will adopt consists in promoting $\xi_0$ to a field, $\xi_0(x)=-\pi(x)$ and requiring the following gauge transformation rule: $\pi(x)\rightarrow \pi(x)-\xi_0(x)$ on $\pi$.  With this assumption in place, 
the above action is invariant under full spacetime diffeomorphisms. The  scalar degree of freedom $\pi$ makes its 
appearance in  the time dependence of the $A(t+\pi),B(t+\pi)$ coefficients and in the transformed metric.
This procedure is essentially the same as the one of standard gauge theory: a Goldstone boson which transforms non-linearly under the gauge transformation provides the longitudinal component of  a massive   gauge boson. For high enough energies,  
the Goldstone  becomes the only relevant degree of freedom. This is the so-called \textit{equivalence} theorem. The same is true for our case: for sufficiently high energy
 the mixing with gravity becomes irrelevant and the scalar $\pi$ becomes the only relevant mode in the dynamics (decoupling regime).  One then needs to identify the scale of the energy above which this approximation holds (on the other side of the energy range one always keeps in mind the upper energy threshold  that comes with the use of effective theory). This procedure puts some bounds on the values of some of the coefficients that drive quadratic operators in the action \cite{eft08,b}. We will not be concerned with these issues because most of the interaction we will be analyzing start at third order in perturbations. It suffices here to say that, since one is concerned with correlators just after horizon crossing, the decoupling procedure works as long as the decoupling energy is smaller than the Hubble rate $H$.\\
\noindent From now on we will work in the decoupling regime. In considering  the terms of Eq.~(\ref{genspacei}), we will therefore use only the unperturbed entries of the metric tensor. To write  the effective Lagrangian up to third order, we  start from  Eq.~(\ref{genspacei}) and follow the algorithm given in \cite{eft08}. Fluctuations are encoded in the $F^{(n)}$ terms. In order to be as general as possible, one must also include all possible 
contributions coming from extrinsic curvature $K_{\mu\nu}$ terms. \\
\noindent Following the procedured outlined above, the third  and fourth-order Lagrangian is obtained. We will use the third-order expression to calculate the contribution to the trispectrum of curvature perturbations arising from the scalar exchange diagram \cite{b}:
\begin{eqnarray}
\fl  S_3&=&\int d^4 x \sqrt{-g}\left[ M_{\rm Pl}^{2}\dot H (\partial_{\mu} \pi)^2 
+ M_2(t)^4\left(2{\dot\pi}^2 -2\dot\pi \frac{(\partial_i \pi)^2}{a^2}\right) -\frac{4}{3}M_3(t)^4{\dot\pi}^3 
\right.\nonumber\\
\fl &-& \frac{\bar M_1(t)^3}{2}\left(\frac{-2 H (\partial_i \pi)^2 }{a^2} +\frac{(\partial_i \pi)^2 \partial_j^2 \pi}{a^4}  \right) 
-\frac{\bar M_2(t)^2}{2} \Big( \frac{(\partial_i^2\pi)(\partial_j^2\pi)   +H (\partial_i^2\pi) (\partial_j \pi)^2}{a^4}
\nonumber\\
\fl    &+&2\frac{\dot\pi \partial_{i}^2 \partial_j \pi \partial_j \pi}{a^4} \Big) 
-\frac{\bar M_3(t)^2}{2} \left( \frac{ (\partial_i^2\pi)(\partial_j^2\pi) +2H(\partial_i \pi)^2\partial_i^2 \pi + 2 \dot\pi \partial_{i  j}^{2}\pi \partial_j \pi  }{a^4} \right)    \nonumber\\
\fl  &-&\frac{2}{3}\bar M_4(t)^3 \frac{1}{a^2}{\dot\pi}^2 \partial_i^2\pi +  \frac{\bar M_5(t)^2}{3}  \frac{\dot\pi}{a^4}(\partial_i^2\pi)^2 
+\frac{\bar M_6(t)^2}{3}  \frac{\dot\pi}{a^4}(\partial_{ij}\pi)^2 -\frac{\bar M_7(t)}{3!} \frac{(\partial_i^2 \pi)^3}{a^6}
\nonumber\\
\fl & -&\left. \frac{\bar M_8(t)}{3!}  \frac{\partial_i^2 \pi}{a^6}(\partial_{jk}\pi)^2-\frac{\bar M_9(t)}{3!}\frac{1}{a^6}\partial_{ij} \pi\partial_{jk} \pi\partial_{ki}\pi
\right].\label{action3}
\end{eqnarray}

The above action and its extension at higher perturbative orders covers many inflationary theories, providing a unifying perspective  which is hard to obtain without an effective approach. Indeed, by switching on and off a single ${\bf M}_n=M_n,\bar M_n$ coefficient one has control over all corresponding operators in the action (the bar on the $\bar M_n$'s signals that these are curvature-generated terms). The hope is to be able to identify distinctive features for as many as possible different combinations of the ${\bf M}_n$'s in the form of specific patterns they produce in the shapes of the various correlators of curvature perturbations. The degeneracies among the results for different inflationary mechanisms that will inevitably arise might be removed by a joint analysis of the different n-point functions, starting with the bispectrum, the trispectrum, loop corrections to the power spectrum and so on.
Let us briefly go through some of the main features of the third order effective action above. All the comments can be straightforwardly extended to the fourth-order expression as well.

\begin{itemize}
\item  Consider only the quadratic terms: for $M_{2}=\bar M_{1,2,3}=0$ one recovers the usual quadratic Lagrangian for the fluctuations, with  sound speed  $c_s^2 =1$. Switching on $M_{2}$ corresponds to allowing models with sound speed smaller than unity, $1/c_s^2 =1- 2M_2^4/(M_{\rm Pl}^2 \dot H)$, which are often linked to a high level of primordial non-Gaussianity \cite{chen-bis,eft08} as for DBI inflation. Further allowing for a non-zero $\bar{M}_{2,3}$ in the de Sitter limit,  one recovers  Ghost Inflation \cite{ghost}. Similarly, having all $\bar M$'s set to zero, and going to third  and higher order with the $M$'s,   one retrieves the interactions characterizing DBI inflation \cite{DBI,chen-bis, chen-tris} and K-inflation \cite{mukh1, mukh2} theories and others. \footnote{See \cite{sarkar, pajer} for some recently introduced examples that require, in order to be recovered in the effective field theory approach, that one relaxes the  implicit assumption of a shift symmetry for the scalar $\pi$.}

\item The action in Eq.~(\ref{action3}) contains in principle additional terms but, being interested in those generating large non-Gaussianities, a selection has been made. Specifically, at every order in fluctuations and for each ${\bf M_n}$ coefficient, only
leading terms are considered. One starts from the realization that, even for the most generic quadratic action the following estimates hold at horizon crossing \cite{b}:
\be
\dot\pi \sim H \pi, \qquad \nabla \pi \sim    \frac{H}{\sqrt{\alpha_0 + \sqrt{\alpha_0^2 + 8 \beta_0}}}\,\, \pi \label{est} \equiv H/ \tilde{c}_s \, \, \pi
\ee
where $\tilde{c}_s$ is a sort of generalized speed of sound.\\ Also, as mentioned before, in the action  the scalar appears only trough its derivatives. When faced with a given ${\bf M}_n$  multiplying terms of the same perturbative order with a given number of derivatives one therefore knows that the leading term will be the one with the most spatial derivatives.\\
There is also the  comparison between the same perturbative order but different ${\bf M}_n$ terms to be made. All non-zero coefficients  in front of the various operators might be assumed to be of the same order \cite{ghost}; interestingly, in \cite{ssz05}, employing renormalization and unitarity arguments, a natural (relative) value of $1/c_s^2$  was obtained for the $M_2^4, M_3^4$ coefficients. In general, we shall not restrict ourselves to these situations. Considering theories with a speed of sound different from unity and allowing for ${\bf M}_n$'s of different orders greatly increases the number of viable terms for large non-Gaussianities. Let us consider an illustrative example. Take the interaction terms
\[
 -2 M_2(t)^4\dot\pi \frac{(\partial_i \pi)^2}{a^2}\,\,; \qquad \qquad  -\frac{\bar M_7(t)}{3!} \frac{(\partial_i^2 \pi)^3}{a^6}
\]
At horizon crossing, the region from which we expect the main contribution to n-point functions, the comparison reads like $M_2^4 H H^2/ \tilde{c}_s^2 \leftrightarrow \bar M_7 H^6/ \tilde{c}_s^6 $. In a Lorentz invariant theory with coefficients of the same order the first term would clearly prevail. Allowing $ \tilde{c}_s \ll 1$ makes the comparison less obvious and an $\bar M_7\gg M_2$ further strengthens this point. Here a word of caution is in order: from simple dimensional analysis a term with ever increasing spatial derivatives will have an ${\bf M}_n$ coefficient with smaller and smaller exponent (counterbalanced by an higher exponent for $H$) and must eventually be subleading with respect to the contributions with fewer derivatives. This is because in the effective theory approach one is roughly making an $H/M$ expansion (with $M$ being the scale of the underlying theory) and, although one can fully employ the freedom to have $ {\bf M}_n$'s of different size up to some perturbative order in order to resuscitate interesting contributions to the correlators,  the ${\bf M}_n$-driven contribution must eventually (from some $n$ on) cease to be relevant.
\end{itemize}
Employing the same calculational algorythm first introduced in \cite{eft08} and used in writing the complete third order action above, we obtain the most general fourth-order action in this set up:

\[
\fl \mathcal{S}_4=\int d^4 x \sqrt{-g}\left[ 
 \frac{1}{2!}M_2(t)^4 \frac{(\partial_i \pi)^4}{a^4} +2 M_3(t)^4 \frac{{\dot\pi}^2 (\partial_i \pi)^2}{a^2} 
+\frac{2}{3}M_4(t)^4 {\dot\pi}^4 
\right.
\]

\[ \left.  
\fl   - \frac{\bar M_1(t)^3}{4}\left(\frac{H(\partial_i \pi)^4}{a^4}-\frac{2 \dot \pi(\partial_i \pi)^2\partial_j^2 \pi}{a^4}   \right)
-\frac{\bar M_2(t)^2}{2} \left(\frac{(\partial_j \pi)^2(\partial_i^2 \pi)^2}{a^6} +\frac {2  \partial_k^2 \pi \partial_i \pi \partial_{ij} \pi \partial_j\pi}{a^6} \right) 
\right. 
\]

\[ 
\left.
\fl -\frac{\bar M_3(t)^2}{2} \left(  \frac{ (\partial_{ij}\pi)^2 (\partial_k \pi)^2}{a^6}+\frac{2 \partial_{i} \pi \partial_{ij} \pi\partial_{jk} \pi\partial_{k} \pi
}{a^6} \right)+\frac{2}{3}\bar M_4(t)^3 \,\, \frac{ \dot\pi (\partial_i \pi)^2 \partial_j^2 \pi}{a^4}
\right.
\]

\[
\left.
\fl -\frac{\bar M_6(t)^2}{3!} \frac{(\partial_k \pi)^2(\partial_{ij} \pi)^2}{a^6} -\frac{\bar M_7(t)}{3!}\left(\frac{3}{2}\frac{(\partial_i^2 \pi)^2 H(\partial_j \pi)^2}{a^6} +\frac{6\,\dot\pi \partial_k^2 \pi (\partial_j\partial_i^2 \pi) \partial_j\pi}{a^6}   \right)
\right.
\]

\bea
\left. 
\fl -\frac{\bar M_8(t)}{3!} \left( \frac{H (\partial_i \pi)^2 (\partial_j^2 \pi)^2}{a^6}+ \frac{H (\partial_i \pi)^2 (\partial_{jk} \pi)^2 }{2a^6}- \frac{2\, H \partial_k^2 \pi \partial_i \pi \partial_{ij} \pi \partial_j \pi}{a^6} + \frac{2\dot \pi \partial_k^2 \pi \partial_i^2 \partial_j \pi \partial_j \pi}{a^6}
\right. \right. \nonumber
\eea
\bea
\left. \left.
 \qquad \qquad + \frac{2\dot \pi \partial_k^2 \partial_i \pi \partial_{ij}\pi \partial_j \pi}{a^6}  +  \frac{2\dot \pi \partial_{ij} \partial_{ijk} \partial_k \pi}{a^6} \right) - \frac{\bar M_5(t)^2}{3!}  \frac{(\partial_i \pi)^2(\partial_j^2 \pi)^2}{a^6}
\right. \nonumber
\eea
\bea
\fl -\frac{\bar M_9(t)}{2}\left( {\frac{ H \partial_k^2 \pi (\partial_{ij}\pi)^2}{2 a^6}-\frac{ H \partial_{i} \pi \partial_{ij}\pi \partial_{jk}\pi \partial_{k}\pi }{a^6}+\frac{ \dot \pi  \partial_{ij}\pi \partial_{ijk} \pi \partial_{k}\pi }{a^6}+\frac{ \dot \pi \partial_i^2 \partial_j \pi \partial_{jk} \pi \partial_k \pi }{a^6}}\right)\nonumber
\eea

\[
\left.
\fl +\frac{\bar M_{10}^{3}(t)}{3}\frac{{\dot\pi}^3\partial_i^2 \pi}{a^2} 
-\frac{\bar M_{11}^{2}(t)}{3!}\frac{{\dot\pi}^2 (\partial_i^2 \pi)^2}{a^4} 
-\frac{\bar M_{12}^{2}(t)}{3!}\frac{{\dot\pi}^2 (\partial_{ij}\pi)^2 }{a^4} +\frac{\bar M_{13}(t)}{4!} \frac{2\,\dot\pi}{a^6}(\partial_i^2\pi)^3 
\right.
\]
\[ \left.  
\fl +\frac{\bar M_{14}(t)}{4!} \frac{2\, \dot\pi \partial_k^2\pi(\partial_{ij}\pi)^2}{a^6}  +\frac{\bar M_{15}(t)}{4!}\frac{2\, \dot \pi  \partial_{ij}\pi \partial_{jk}\pi \partial_{ki}\pi }{a^6} -\frac{\bar N_{1}(t)}{4!}\frac{(\partial_i^2 \pi)^4}{a^8} -\frac{\bar N_{2}(t)}{4!}\frac{(\partial_k^2 \pi)^2(\partial_{ij}\pi)^2}{a^8} 
\right.
\]
\bea
\left.
\fl  -\frac{\bar N_{3}(t)}{4!}\frac{\partial_{\rho}^2\pi   \partial_{ij}\pi \partial_{jk}\pi \partial_{ki}\pi}{a^8} -\frac{\bar N_{4}(t)}{4!}\frac{ (\partial_{ij} \pi)^4 }{a^8} -\frac{\bar N_{5}(t)}{4!} \frac{\partial_{ij}\pi \partial_{jk}\pi \partial_{k \rho}\pi \partial_{\rho i}\pi}{a^8}\right] \label{l4}
\eea

Note that, as pointed out in \cite{lh}, starting at fourth order in perturbations, one cannot immediately read off the Hamiltonian from the expression of the Lagrangian, in other words $H = - L$ does not hold here. We use the results one obtains by adopting the correct procedure which was outlined in detail in \cite{lh}.\\
Let us split the interaction Hamiltonian we will be concerned with as $H_{int}= H_3 + H_4$; one can prove that the overall interaction Hamiltonian is then:
\bea
\fl H_{int}= - L_3 - L_4 + \int d^3 x \sqrt{-g}\Big[ \frac{1}{2 M_2^4+H^2 \epsilon  M_P^2-3 H \bar{M}_1^3} \left( \frac{(\partial_i \pi)^4 M_2^8}{a^4} +\frac{4 {\dot \pi}^2 (\partial_i \pi)^2 M_2^4 M_3^4}{a^2} \right. \nonumber \\
\fl \left. +4 {\dot \pi}^4 M_3^8+\frac{(\partial_k \pi)^2 \partial_{i}^2 \partial_j \pi  \partial_j \pi M_2^4 \bar{M}_2^2}{a^6 }+\frac{2 {\dot \pi}^2 \partial_{i}^2 \partial_j \pi  \partial_j \pi M_3^4 \bar{M}_2^2}{a^4} +\frac{(\partial_{i}^2 \partial_j \pi  \partial_j \pi)^2  \bar{M}_2^4}{4 a^8} \right.  \nonumber \\
\fl \left. +\frac{(\partial_k \pi)^2 \partial_i^2 \partial_j \pi \partial_j \pi M_2^4 \bar{M}_3^2}{a^6 }+\frac{2 {\dot \pi}^2  \partial_i^2 \partial_j \pi \partial_j \pi M_3^4 \bar{M}_3^2}{a^4 } +\frac{( \partial_i^2 \partial_j \pi \partial_j \pi)^2 \bar{M}_2^2 \bar{M}_3^2}{2 a^8 } +\frac{(\partial_i^2 \partial_j \pi \partial_j \pi)^2 \bar{M}_3^4}{4 a^8} \right.\nonumber \\
\fl\left.  +\frac{4 {\dot \pi} (\partial_k \pi)^2 \partial_i^2 \pi M_2^4 \bar{M}_4^3}{3 a^4}  +\frac{8 {\dot \pi}^3 {\partial_i^2 \pi} M_3^4 \bar{M}_4^3}{3 a^2}  
+\frac{2 {\dot \pi} {\partial_k^2 \pi} {\partial_i^2 \partial_j \pi} {\partial_j \pi} \bar{M}_2^2 \bar{M}_4^3}{3 a^6} +\frac{2 {\dot \pi} {\partial_k^2 \pi}  {\partial_i^2 \partial_j \pi} {\partial_j \pi} \bar{M}_3^2 \bar{M}_4^3}{3 a^6 }
\right. \nonumber \\
\left. \fl +\frac{4 {\dot \pi}^2 (\partial_k^2 \pi)^2 \bar{M}_4^6}{9 a^4 }-\frac{(\partial_i \pi)^2 (\partial_k^2 \pi)^2 M_2^4 \bar{M}_5^2}{3 a^6 }-\frac{2 {\dot \pi}^2 (\partial_i^2 \pi)^2 M_3^4 \bar{M}_5^2}{3 a^4 } -\frac{(\partial_k^2 \pi)^2  \partial_i^2 \partial_j \pi \partial_j \pi \bar{M}_2^2 \bar{M}_5^2}{6 a^8} \right.  \nonumber \\
\fl \left.  -\frac{(\partial_k^2 \pi)^2 \partial_j \pi \partial_j \pi \bar{M}_3^2 \bar{M}_5^2}{6 a^8 }-\frac{2 {\dot \pi} (\partial_k^2 \pi)^3 \bar{M}_4^3 \bar{M}_5^2}{9 a^6 }  +\frac{(\partial_k^2 \pi)^4 \bar{M}_5^4}{36 a^8}-\frac{(\partial_k \pi)^2 (\partial_{ij}\pi)^2 M_2^4 \bar{M}_6^2}{3 a^6 } \right.  \nonumber \\
\fl \left. -\frac{2 {\dot \pi}^2 (\partial_{ij}\pi)^2 M_3^4 \bar{M}_6^2}{3 a^4 } -\frac{ (\partial_{kl}\pi)^2 \partial_i^2 \partial_j \pi \partial_j \pi \bar{M}_2^2 \bar{M}_6^2}{6 a^8 \left(2 M_2^4+H^2 \epsilon  M_P^2-3 H \bar{M}_1^3\right)}-\frac{(\partial_{kl}\pi)^2 \partial_i^2 \partial_j \pi \partial_j \pi \bar{M}_3^2 \bar{M}_6^2}{6 a^8 \left(2 M_2^4+H^2 \epsilon  M_P^2-3 H \bar{M}_1^3\right)} \right. \nonumber \\
\fl \left. -\frac{2 \dot \pi {\partial_k^2 \pi} (\partial_{ij}\pi)^2 \bar{M}_4^3 \bar{M}_6^2}{9 a^6}  +\frac{(\partial_k^2 \pi)^2 (\partial_{ij}\pi)^2 \bar{M}_5^2 \bar{M}_6^2}{18 a^8 }+\frac{(\partial_{ij}\pi)^4 \bar{M}_6^4}{36 a^8 } \right) \Big]  \nonumber \\
\label{h4}
\eea 
where the above terms besides $-(L_3 + L_{4})$ are all at fourth order in perturbations.\\
\section{Symmetries}
Having written the complete Hamiltonian, we now proceed to calculate the four-point function contributions arising from interaction terms at third and fourth order. We employ here the IN-IN formalism \cite{in-in1, in-in2, in-in3, w-qccc} and conveniently split the contributions to the four-point function as the ones arising from terms that make up the contact interaction diagram and the ones that generate the scalar exchange diagram as in the figure below.


\begin{center}
\fcolorbox{white}{white}{
  \begin{picture}(370,66) (47,-47)
    \SetWidth{1.0}
    \SetColor{Black}
    \Line(48,-46)(80,-14)
    \Line(80,-14)(48,18)
    \Line(80,-14)(128,-14)
    \Line(128,-14)(160,18)
    \Line(128,-14)(160,-46)
    \Line(352,-46)(416,18)
    \Line(352,18)(416,-46)
 \Text(8,-65)[lb]{\small{\Black{\textbf{Figure A}: On the left, the scalar exchange diagram. Contact interaction diagram on the right.}}}
  \end{picture}
}
\end{center}

\vspace{4mm}
\noindent It is useful at this stage to offer some comments on the calculations we are going to present. As mentioned, the literature already contains a thorough analysis of the trispectra for \textit{general single-field inflation} models, see for example \cite{chen-tris}. Work on the four-point function for ghost inflationary models has recently been presented \cite{huang, muko}. Our starting point, being based on a comprehensive effective theory, clearly encompasses all these models. Working with the effective Hamiltonian above translates into many immediate advantages as listed before but, on the other hand, in calculating the resulting four point function, one faces a substantial number of terms and it is therefore natural to look for some ordering principle which would single out some contributions to the trispectrum as the leading ones and allow us to concentrate on them only. In this context employing a symmetry for the whole theory can prove very useful. Indeed in \cite{4pt,muko} the authors consider only those allowed by a particular (approximate in \cite{4pt}) symmetry of the action, respectively:
\be
{\bf S1}:\,\,\, \pi \rightarrow -\pi\,\, ; \qquad \qquad {\bf S2}:\,\,\, \pi \rightarrow - \pi\quad\,\,\textit{and} \,\,  \quad t \rightarrow -t\,\, . \label{symm}
\ee
We plan here to employ our general effective theory to show that, allowing some freedom on the ${\bf M}_n$ coefficients that modulate the various terms in the third and fourth order action, within each one of the two distinct and quite restrictive symmetry requirements above there are novel curvature-generated terms in the action that should not be disregarded as negligible and that, furthermore, show some distinctive features in the shapes of the trispectrum. We will also describe terms allowed by both the symmetries in Eq.~(\ref{symm}) combined. Of course, one need not employ  symmetries to switch on or off any specific operator in the action. Most of the contributions are indeed freely adjustable by the correspondent $M_n$ coefficient, a procedure which is, in principle, legitimate since the underlying theory is unknown. We choose here to restrict ourselves to considering only symmetry-abiding terms. Let us comment on each one of the symmetries.\\ {\bf S1} is built upon the following considerations. Often the same ${\bf M}_n$ coefficients multiply terms of different perturbative orders; consequently the amplitude  $f_{NL}$ of the 3-point function will be related to the amplitude of higher order correlators, notably to $\tau_{NL}$, the amplitude for the four-point function. Whenever the leading part of the trispectrum is generated by these types of ${\bf M}_n$'s one can estimate that for its effect to be observable $\tau_{NL}$ has to be five orders of magnitude larger than $f_{NL}$ \cite{4pt}, which leaves little room for feasible models. On the other hand, one quickly realizes those ${\bf M}_n$'s whose first term starts only at the fourth perturbative order ($M_4, \bar M_{10}...$in Eq.~(\ref{l4})) are not plagued by this problem. This then represents a natural way to obtain inflationary models which allow a large, detectable trispectrum untied to the interactions which make up the bispectrum (which might well be small now).\footnote{One needs also to check that the interactions driven by coefficients that multiply also third order fluctuations do not become important in the form radiative corrections to the bispectrum. This check is done in \cite{4pt} and ensures that loop corrections of those terms are not relevant.} Indeed, in \cite{4pt} the authors investigate on the size of all the interactions driven by the $M_4, \bar M_{11}, \bar N_1$ \footnote{In the same spirit of the analysis done in \cite{b} for all curvature-generated terms at third order, the authors of \cite{4pt} consider in the v2 of their paper some extrinsic-curvature terms generated at fourth order. They also comment on their importance in near de Sitter limit and their conclusions apply to our $\bar M_{11}, \bar N_1$ parameters.} coefficients in Eq.~(\ref{l4}) and show that the leading interactions driven by these parameters  are all consistent with the $\pi \rightarrow - \pi$ prescription and are expected to give a comparable signal \footnote{It would be interesting to understand to what kind of models, in terms of the fundamental scalar field, the simple resulting effective Lagrangian corresponds in this case. }. By construction then, the terms in the interaction Hamiltonian that are going to contribute to the trispectrum and be consistent with the reasoning that inspired the {\bf S1} symmetry are only some of the ones that will make up the contact interaction diagram, namely those whose lowest order interaction is already at fourth order. This limits us to the contributions regulated by the following coefficients: $M_4, \bar M_{10}..\bar M_{15}, \bar N_{1}..\bar N_{5}$.\\
{\bf S2} symmetry, on the other hand, does not prohibit third order interactions, indeed in \cite{muko} the interaction $\dot \pi (\nabla \pi)^2$ is considered and, by inspection of Eq.~(\ref{action3}), one can see that also other terms are allowed, the one regulated by $\bar M_5$ and, notably, the $\bar M_6\, \dot \pi (\partial_{ij}\pi)^2/a^4$ term. The $\bar M_6$-driven term is particularly interesting because its contribution to the bispectrum calculations of \cite{b} generates an interesting flat shape. The scalar exchange diagram will then be built out of the third order {\bf S2}-obeying terms in the action. In particular, inspired by previous findings, we are going to give a detailed account of the $\bar M_6$ contribution.\\
If both {\bf S1} \textit{and} {\bf S2} are to be enforced one must also exclude from the list of {\bf S1}-abiding interactions the ones multiplied by $\bar M_{10},\bar M_{13},\bar M_{14},\bar M_{15}$. A more clear picture of the situation concerning the various symmetries is presented in \textit{Table 1} below.\\

\begin{center}
\textbf{Table 1}\\
\vspace{5mm}
\begin{tabular}{| l || l | l | l | l | l| l | l| l| l| l| l|l|l| }
\hline			
       Coefficients   & $M_2$ & $M_3$ & $M_4$ & $\bar M_1$ & $\bar M_2$ & $\bar M_3$ & $\bar M_4$ & $\bar M_5$ & $\bar M_6$ & $\bar M_7$ & $\bar M_8$ & $\bar M_9$  \\ \hline 
  ${\bf S1}$ & X & X & \checkmark & X & X & X & X & X & X & X & X &X    \\ \hline
  ${\bf S2}$  & \checkmark &\checkmark  & \checkmark & X & X & X & X & \checkmark  & \checkmark & X & X   & X \\\hline
  Coefficients & $\bar M_{10}$ & $\bar M_{11}$ & $\bar M_{12}$ & $\bar M_{13}$ & $\bar M_{14}$ & $\bar M_{15}$ & $\bar N_1$ & $\bar N_2$ & $\bar N_3$ & $\bar N_4$ & $\bar N_5$ & $/$    \\ \hline 
  ${\bf S1}$  & \checkmark  & \checkmark & \checkmark & \checkmark & \checkmark & \checkmark & \checkmark & \checkmark & \checkmark & \checkmark &\checkmark &    \\ \hline
  ${\bf S2}$  & X & \checkmark  & \checkmark  & X & X & X & \checkmark & \checkmark & \checkmark & \checkmark & \checkmark  &  \\\hline
    
\end{tabular}
\end{center}
{\small { \bf  The Coefficients marked with `` \checkmark '' in correspondence of a given symmetry {\bf S} are {\bf S}-invariant, those marked with ``X'' violate the {\bf S} symmetry.  }}
\\
\vspace{4mm}

Note that each ${\bf M}_n$ coefficient might multiply many interactions at each perturbative orders and therefore we mark the coefficient as invariant under a symmetry when all the \textit{leading} interactions it multiplies are invariant under {\bf S1} or {\bf S2}. Determing the properties of the coefficients in the second row requires no effort, as one can easily verify these ${\bf M}_n$'s first appear in the action as multipliers of fourth-order terms. Things are less linear with the coefficients in the first row (except for $M_4$) as they appear at fourth order both multiplying bare interaction terms and multiplying other coefficients as well as interaction terms (for an example of the latter case see the terms written explicitly in Eq.~(\ref{h4})). They also appear at third and some also at second order in perturbations. One then must carefully check that, given a particular coefficient ${\bf M}_n$, in none of the interactions it multiplies at any order the leading terms violate the symmetry. For $M_2, M_3, \bar M_5, \bar M_6$ in the first row one can verify after some checks that these terms all parametrize indeed approximately invariant interactions upon requiring the coefficient $\bar M_4^3$ to be much smaller than the typical ${\bf M}_n$ such as  $M_2..\bar M_6$. This is because in the fourth-order Hamiltonian in Eq.~(\ref{h4}) there are terms of the form

\be
\propto \frac{1}{M^4}\,\, \bar M_4^3\times \{M_2^4, M_3^4, \bar M_5^2, \bar M_6^2 \}\times ({\bf S2}-\textit{violating\,\, interaction})
\ee
which one then assumes to be subleading. We stress this point because it emerges clearly and naturally in the effective theory approach.\\

\section{Trispectrum}
\subsection{IN-IN Formalism}
We are going to employ the IN-IN formalism to calculate the four point function of curvature perturbation. The most general and compact expression for such a quantity is:

\bea
\fl \langle \Omega|\zeta_{k1}\zeta_{k2}\zeta_{k3}\zeta_{k4}(t)|\Omega \rangle = \langle 0|\bar T \{e^{i\int_{-\infty}^{t_0} d^3 x dt^{'} \mathcal{H}(x)}\} \zeta_{k1}\zeta_{k2}\zeta_{k3}\zeta_{k4}(t)\,  T\{ e^{-i\int_{-\infty}^{t_0} d^3 x^{'} dt^{''} \mathcal{H}(x)} \}|0\rangle \nonumber\\ , \label{ttb}
\eea
where $\bar T$ and $T$ indicate respectively anti-time order and time order operations,  $|0\rangle$ and $|\Omega\rangle$ stand for the vacuum of the free and interacting theory.\\
Expanding both the exponentials in Eq.~(\ref{ttb}), we single out the first non vanishing terms that will contribute to the scalar exchange and contact interaction diagrams.
\bea
 \langle \Omega|\zeta_{k1}\zeta_{k2}\zeta_{k3}\zeta_{k4}(t)|\Omega\rangle = \nonumber\\ \langle 0|\bar T \{i\int_{-\infty}^{t_0} d^3 x dt^{'} \mathcal{H}_3(x)\} \zeta_{k1}\zeta_{k2}\zeta_{k3}\zeta_{k4}(t)\,  T\{ -i\int_{-\infty}^{t_0} d^3 x^{'} dt^{''} \mathcal{H}_3(x^{'}) \}|0\rangle \nonumber\\
 +\langle 0|\bar T \{\frac{i^2}{2}\int{ \int{ d^3 x \,dt^{'}\, d^3 x^{'}\, dt^{''} \mathcal{H}_3(x) \mathcal{H}_3(x^{'})}}\} \zeta_{k1}\zeta_{k2}\zeta_{k3}\zeta_{k4}(t)|0 \rangle \nonumber\\
 +\langle 0|\zeta_{k1} \zeta_{k2}\zeta_{k3}\zeta_{k4}(t) T \{\frac{(-i)^2}{2}\int_{-\infty}^{t_0}{ \int_{-\infty}^{t_0}{ d^3 x \,dt^{'}\, d^3 x^{'}\, dt^{''} \mathcal{H}_3(x) \mathcal{H}_3(x^{'})}}\}|0\rangle
\nonumber\\
 +\langle 0|\bar T \{i\int_{-\infty}^{t_0}{d^3 x \,dt\, \mathcal{H}_4(x)}\} \zeta_{k1}\zeta_{k2}\zeta_{k3}\zeta_{k4}(t)|0\rangle \nonumber\\
 +\langle 0|\zeta_{k1} \zeta_{k2}\zeta_{k3}\zeta_{k4}(t) T \{-i\int_{-\infty}^{t_0}{ d^3 x^{'}\, dt^{''}  \mathcal{H}_4(x^{'})}\}|0\rangle +...\label{z4}
\eea
where  $\mathcal{H}_3,\mathcal{H}_4 $ are the third and fourth-order Hamiltonian in the interaction picture. The latter two terms make up the contact interaction diagram, the rest is responsible for the scalar exchange. Let us also remind the reader that the gauge invariant observable $\zeta$ is, at first approximation, linearly related to the scalar $\pi$ via $\zeta=-H \pi$. Also, already at this stage one can see that the result of the four point function is going to depend on  six variables. All wavefunctions, once in Fourier space, depend only on the magnitude of their momenta. There are at most ten fields involved in the contractions, eight of which will always depend on the magnitude of the four external momenta ($k_1,k_2,k_3,k_4$). We are left with one last contraction between two fields depending on the magnitude of one vector which, by construction, is going to be the sum of two external momenta. It turns out that, employing the overall momentum conservation, two variables are sufficient to describe any of these linear combinations, we choose $(k_{12} \equiv |\vec k_1+ \vec k_2|,k_{14} \equiv |\vec k_1+ \vec k_4|)$, giving  a total of six variables. As clear from above, the $\bar M_6$-driven third-order interaction we are going to consider further depends on scalar products between the various momenta but, as one can easily verify, these can all be fully specified by using the six variables introduced above. All the variables we will employ are represented in the figure below.

\begin{center}
\fcolorbox{white}{white}{
  \begin{picture}(276,196) (191,-43)
    \SetWidth{1.0}
    \SetColor{Black}
    \Line[arrow,arrowpos=0.5,arrowlength=5,arrowwidth=2,arrowinset=0.2](192,52)(400,4)
    \Line[arrow,arrowpos=0.5,arrowlength=5,arrowwidth=2,arrowinset=0.2](400,4)(432,84)
    \Line[arrow,arrowpos=0.5,arrowlength=5,arrowwidth=2,arrowinset=0.2](432,84)(256,148)
    \Line[arrow,arrowpos=0.5,arrowlength=5,arrowwidth=2,arrowinset=0.2](256,148)(192,52)
    \Line[arrow,arrowpos=0.5,arrowlength=5,arrowwidth=2,arrowinset=0.2](256,148)(400,4)
    \Line[dash,dashsize=2,arrow,arrowpos=0.5,arrowlength=5,arrowwidth=2,arrowinset=0.2](192,52)(432,84)
    \Text(272,14)[lb]{\Large{\Black{$k_1$}}}
    \Text(428,36)[lb]{\Large{\Black{$k_2$}}}
    \Text(360,122)[lb]{\Large{\Black{$k_3$}}}
    \Text(208,116)[lb]{\Large{\Black{$k_4$}}}
    \Text(276,94)[lb]{\Large{\Black{$k_{14}$}}}
    \Text(368,57)[lb]{\Large{\Black{$k_{12}$}}}


    \Arc[clock](366.455,20.679)(15.531,-158.551,-279.431)
    \Text(348,32)[lb]{\small{\Black{$\alpha$}}}
    \Arc[clock](397.214,19.929)(13.258,175.365,27.255)
    \Text(397,22)[lb]{\small{\Black{$\beta$}}}
    \Arc[dash,dashsize=10,clock](392.783,7.705)(33.382,169.13,54.852)
    \Text(366,16)[lb]{\small{\Black{$\gamma$}}}
  \Text(108,-12)[lb]{\small {\bf Figure B:} the regular tetrahedron described by the four external momenta and $k_{12}, k_{14}$}
   \end{picture}
}
\end{center}
\noindent In order to get a tetrahedron as the one in Fig~B one must enforce the following inequalities:
\be
 cos(\alpha -\beta) \ge cos(\gamma)\ge cos(\alpha+\beta)
\ee
with 
\be
\fl cos(\alpha)= \frac{k_1^2+k_{14}^2-k_4^2}{2k_1 k_{14}}\,; \quad cos(\beta)= \frac{k_2^2+k_{14}^2-k_3^2}{2k_2 k_{14}}\,;\quad cos(\gamma)= \frac{k_1^2+k_{2}^2-k_{12}^2}{2k_1 k_{2}}  .
\ee
From $-1 \le cos(\alpha,\beta,\gamma) \le 1$ one also obtains the usual  triangles inequalities. Here we single out some of the inequalities which we are going to use in what follows: 
\bea
k_{12}^2 + k_{14}^2 \le 4\, ; \quad \sqrt{1+k_4^2 -2k_4} \le k_{14}\le \sqrt{1+k_4^2 +2k_4}. \label{ineq}
\eea
In order to have a visual intuition and understanding of  the result, once the calculation of the several contributions to the trispectrum is performed  one needs to set up a number of configurations in which four out of the six variables are held fixed. Having more than one configuration also increases one's ability to distinguish the signatures of different interactions. Following \cite{chen-tris}, we adopt the set up described below:
\begin{itemize}
\item \textit{Equilateral configuration}: all the external momenta have the same magnitude $k=k_1=k_2=k_3=k_4$; the two variables left are plotted as $k_{12}/k, k_{14}/k$. Note that when plotting in this configuration we will use the first inequality in Eq.~(\ref{ineq}). Incidentally, this is the only configuration for which exact calculations for the trispectrum in ghost inflation have been presented (see \cite{muko}) so far. Note also that for the equilateral as well as for the other configurations, one conveniently plots the result of the calculations in Eq.~(\ref{z4}) for any specific interaction term multiplied by a factor of $\prod_{i=1}^{4}k_i^3$. It is done also because this factor is generally common to all the contributions and so removing it sharpens the differences between the plots of each interaction term.
\item \textit{Folded configuration}: here one has $k_{12}\rightarrow 0$ as well as $k_1 = k_2$ and $k_3 = k_4$. The second and third inequalities in Eq.~(\ref{ineq}) must be enforced in this case. The variables $k_{14}$ and $k_4$ are the ones plotted in this configuration.
\item \textit{Specialized planar limit configuration}: in this case we have $k_1=k_3=k_{14}$ as well as:
\be
k_{12}=\Big[k_1^2+\frac{k_2 k_4}{2\,k_1^2}\Big(k_2 k_4 + \sqrt{(4k_1^2 -k_2^2)(4k_1^2 -k_4^2)} \,\,   \Big)     \Big]^{1/2}.
\ee
The variables plotted are going to be $k_2/k_1$ and $k_4/k_1$.
\item \textit{Near double squeezed limit configuration}: the tetrahedron is now a planar quadrangle and $k_3 = k_4 = k_{12}$. The region of interest is in particular the one for which $k_3,\,k_4,\,\, k_{12}\rightarrow 0$ where the following relation holds:
\bea
\fl k_2=\frac{\sqrt{k_1^2(-k_{12}^2 +k_3^2+k_4^2)- k_{s1}^2 k_{s2}^2+k_{12}^2 k_{14}^2+k_{12}^2 k_{4}^2+k_{14}^2 k_{4}^2-k_{14}^2 k_{3}^2-k_4^4 +k_{3}^2 k_{4}^2}}{\sqrt{2}k_4} \nonumber\\
\eea

with
\bea
\fl  k_{s1}^2=2 \sqrt{(k_1 k_4 + {\bf k}_1 \cdot {\bf k}_4)(k_1 k_4 - {\bf k}_1 \cdot {\bf k}_4)} \quad  k_{s2}^2= \sqrt{(k_3 k_4 + {\bf k}_3 \cdot {\bf k}_4)(k_3 k_4 - {\bf k}_3 \cdot {\bf k}_4)}. \nonumber \\
\eea
In this case as well the last two inequalities of Eq.~(\ref{ineq}) will be imposed on the variables $k_{14}/k_1$ and $k_4/k_1$. Note that, only in this configuration, what one actually plots is the result of Eq.~(\ref{z4}) times  $\prod_{i=1}^{4}k_i^2$, instead of $\prod_{i=1}^{4}k_i^3$. This is once again done in order to better appreciate the difference among the many interaction terms. 
\end{itemize}
 We now consider  the result for the scalar exchange contribution focusing in particular on an interaction term (the one proportional to $\bar M_6$ in Eq.~\ref{action3} ) case which proved very interesting in plotting the shape of the bispectrum \cite{b}.\\
In all the calculations that follow we use a simplifying assumption which has been verified to hold for 3-point functions and is expected to hold for higher correlators as well \cite{b}. Instead of using the generalized wavefunction which comprises the \textit{general single-field inflation} solution, the Ghost inflation one, etc. as its simplified limits, we employ the usual Hankel function $H_{3/2}(k \tilde{c_s}\tau)$ as a solution to the equation of motion for the quadratic action. The rationale for such a simplification is that, as one can readily verify, the main contribution to higher order correlators comes as usual from the horizon-crossing region and precisely in that region the behaviour of the general solution of \cite{b} (see also \cite{p.s.}) resembles very closely the one of the simpler specific DBI wavefunction. We elaborate further on this fact in \textit{Appendix B} where some examples and comparisons of explicit calculations are provided.\\
\noindent Before moving to the detailed analysis of the shape-functions for several interactions terms in various configurations, let us comment briefly on the amplitudes generically associated with these interaction terms. As noted before, building on the freedom on the ${\bf M}_n$ coefficients allowed by the theory and on the possibility of employing a small speed of sound, $c_s^2 \ll 1$ (the same holds for the parameters which represent the generalization of $c_s^2$, i.e. $\alpha_0$ and $\beta_0$), one can obtain large values for the amplitude associated to each one of the curvature-generated terms we are going to study. This has been quantitatively verified for all the terms of Eq.~(\ref{action3}) in \cite{b}. As an example, consider the $\bar M_6$-driven fourthorder interaction term. To estimate the size of the amplitude associated with a given interaction term one considers its ratio with the quadratic terms of the theory at freezing \cite{eft08}. Applying this prescription to our example one obtains:
\be
\frac{\mathcal{L}_{\bar M_6^2 (\nabla \pi )^2(\nabla^2 \pi)^2}}{\mathcal{L}_2}\sim \frac{\bar M_6^2 \,H^6\,\pi^4/\tilde{c_s}^6 }{M_2^4 \,H^2 \,{\pi}^2}\sim \frac{\bar M_6^2 \,H^2 }{M_2^4}\frac{\zeta^2}{\tilde{c_s}^6}
\ee 
 where the linear relation $\zeta = - H \pi $ has been used; taking $\zeta \sim 10^{-5}$ gives a rough estimate of the size of the non linear corrections.
\subsection{Scalar exchange diagram}
\begin{figure}[hp]
	\includegraphics[scale=0.50]{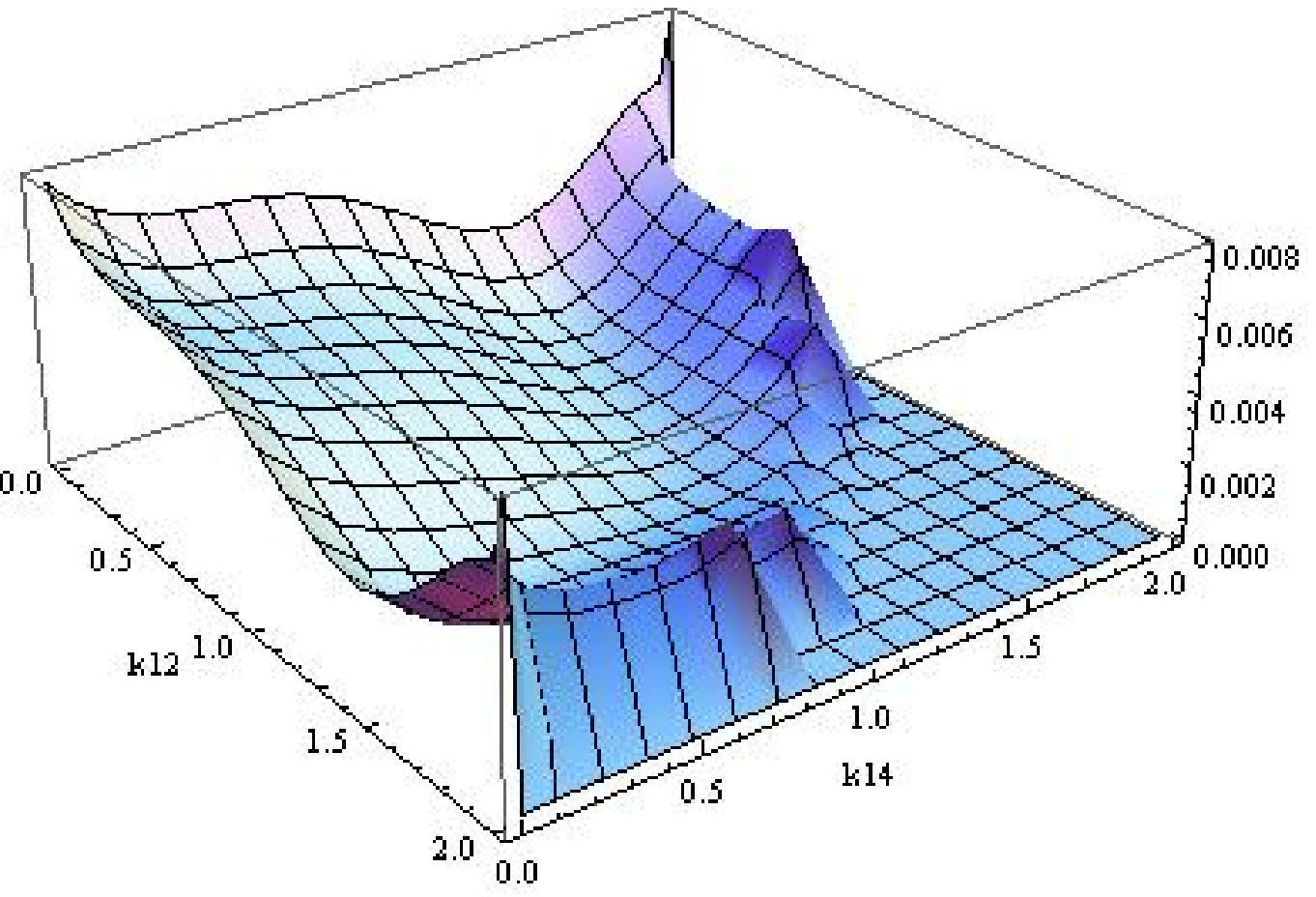}
	\hspace{10mm}
		\includegraphics[scale=0.50]{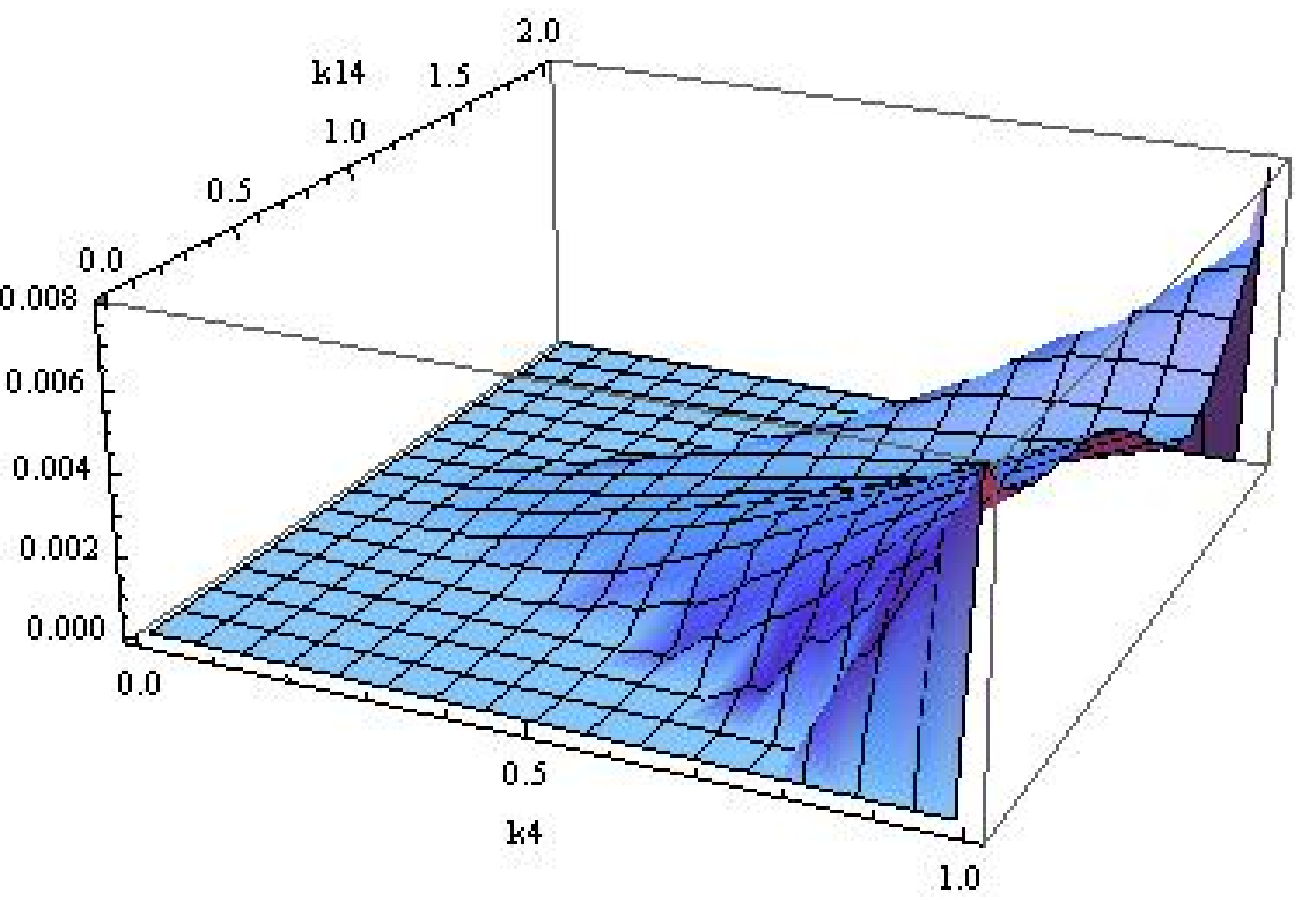}
\caption{The \textit{equilateral} configuration shape is presented on the left from the scalar exchange contribution of the $\bar M_6$-driven interaction. The shape-function is different from the plots presented in \cite{chen-tris}. On the other hand, once a necessary change of variables has been performed, it is qualitatively similar to the shape for the contact interaction diagram which arises from the ghost inflation $(\nabla \pi)^4$ interaction term in \cite{muko}.\\ On the right we plotted our findings for the $\bar M_6$-generated interaction in the \textit{folded} configuration. It very much resembles the ones obtained in \cite{chen-tris} for the scalar exchange diagrams from DBI-like terms, especially from the interaction $\dot \pi (\nabla \pi)^2$ . In all the pictures above and below $k_1$ has ben set equal to unity without loss of generality.}
\label{m61}
\end{figure}

\begin{figure}[hp]
	\includegraphics[scale=0.54]{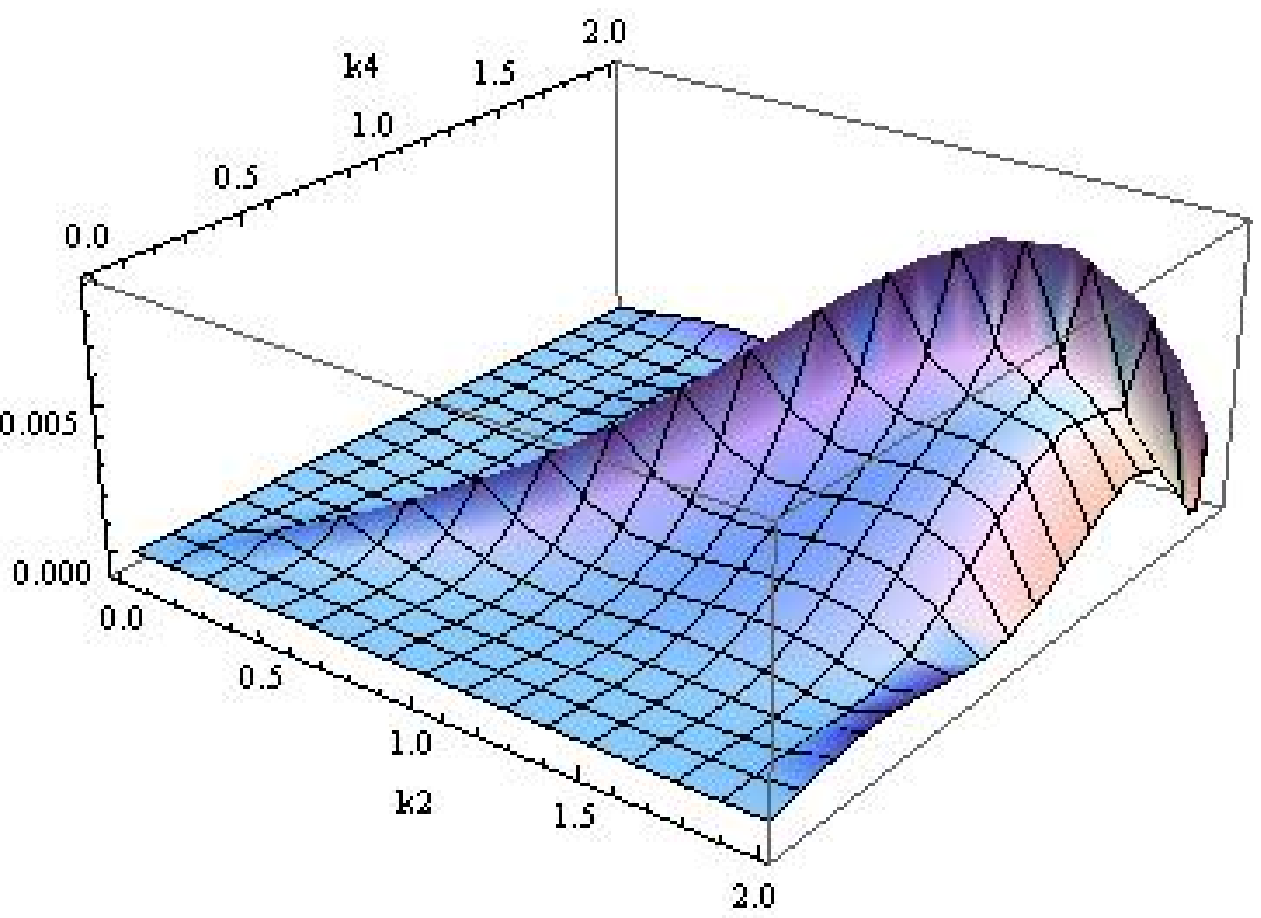}
	\hspace{20mm}
		\includegraphics[scale=0.43]{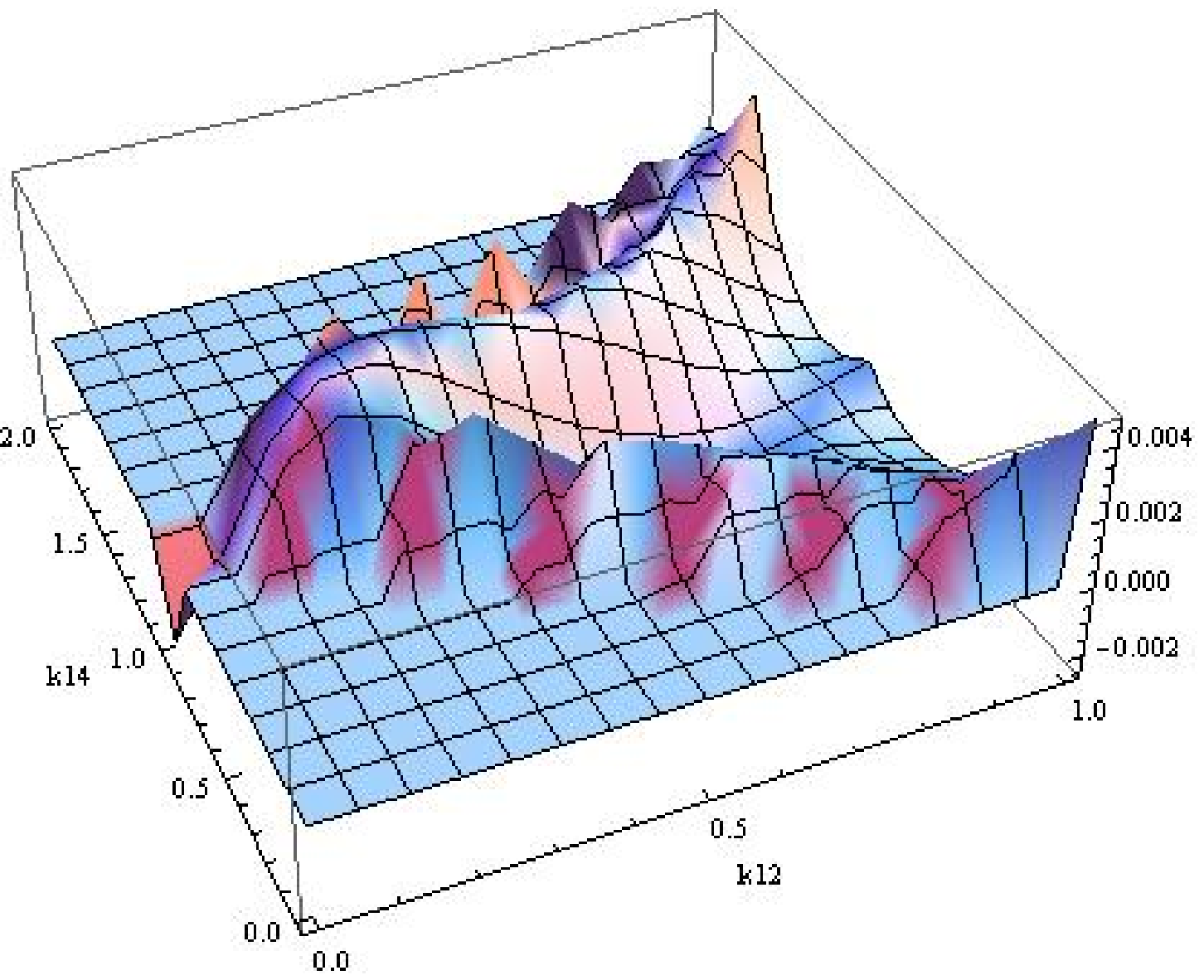}
\caption{On the left the plot obtained for the $\bar M_6$ interaction in the \textit{planar} configuration plot. Note  here some of the interesting features: as $k_2,k_4 \rightarrow 0$ the shape function goes to zero, much like it happens for DBI-generated interactions. As $k_2,k_4 \rightarrow 2$ the shape function reaches values that are negative, albeit slightly so. Looking at the $k_2 = k_4$ line we see that it is convex, rather than concave as found for other interaction types in \cite{chen-tris}\\ On the right the \textit{planar limit double squeezed configuration} is plotted. In the region of interest, namely for $k_{12} \rightarrow 0$, the shape function is non-zero, finite and negative; this again is different than what found in \cite{chen-tris} for a variety of DBI-originated terms.  }
\label{m62}
\end{figure}

Here we are going to consider the interaction term $\bar M_6\,\, \dot \pi (\partial_{ij}\pi)^2/a^4$. Note that, as opposed to the $\bar M_8, \bar M_9$-regulated terms in Eq.~(\ref{action3}) which give a flat shape for the bispectrum much like the $\bar M_6$-driven interaction, this term is actually invariant under the symmetry {\bf S2} while for it to be (approximately) invariant under {\bf S1} one needs to require its $\bar M_6$ coefficient to be much smaller than $M_4$ which would in turn make its signal undetectable.
We now write more explicitly the contribution of the $\bar M_6$-driven third order interaction to the scalar exchange diagram. For all the details of the calculation, including contractions, we refer the reader to \textit{Appendix A}. Consider here just one particular contraction of the fields, the sample contribution we are after looks like the following: 
\bea
\fl <\pi_{k_1} \pi_{k_2}\pi_{k_3}\pi_{k_4}>^{s.e.}_{\bar M_6}= \nonumber \sum_{{\small \textit{all contractions}}}
\eea
\bea
\fl \pi^{*}_{k_1}\pi^{*}_{k_2}\pi_{k_3}\pi_{k_4}(0)\,\int_{-\infty}^{t\rightarrow \,0}{dt_1  a^3 \dot \pi_{k_{12}} \pi_{k_1}\pi_{k_2}({\bf k}_1\cdot {\bf k}_2)^2 /a^4}\int_{-\infty}^{t\rightarrow \,0}{dt_2  a^3 \dot \pi^{*}_{k_{12}} \pi^{*}_{k_3}\pi^{*}_{k_4} ({\bf k}_3\cdot {\bf k}_4)^2 /a^4}  \nonumber\\
\eea
\bea
\fl - 2\mathcal{R}_e \Big[ \pi^{*}_{k_1}\pi^{*}_{k_2}\pi^{*}_{k_3}\pi^{*}_{k_4}\,\int_{-\infty}^{t\rightarrow \,0}{dt_1   \frac{a^3}{a^4}  \dot \pi^{*}_{k_{12}} \pi_{k_1}\pi_{k_2}(t_1)({\bf k}_1\cdot {\bf k}_2)^2 }\int_{-\infty}^{t_1}{dt_2  \frac{a^3}{a^4} \dot \pi_{k_{12}} \pi_{k_3}\pi_{k_4}(t_2) ({\bf k}_3\cdot {\bf k}_4)^2 }   \Big], \nonumber \\ \label{calc}
\eea
where an overall momentum conservation delta and a factor of $(2\pi)^3$ have been omitted for simplicity.\\In Fig. \ref{m61}, \ref{m62} below we plot what one obtains by summing over all contractions, accounting for the symmetry factors of the vertices and plotting the result.\\ For the sake of comparison we often make reference to the shapes obtained in \cite{chen-tris} and \cite{muko}. In the former work the so called local trispectrum is also plotted and compared with findings for \textit{general single-field inflation} models; in the latter one shapes for the trispectrum of ghost inflation are presented in the equilateral configuration only. As mentioned, our starting Lagrangian comprises both these inflationary models; we decided to concentrate on plotting the novel curvature-generated terms that are invariant under {\bf S1}, {\bf S2} or both. 

\newpage

Overall we see that performing the shape analysis for this {\bf S2}-abiding term we are able to find some distinctive features with respect to the DBI-generated contributions.  This is quite interesting also given the fact that the very same $\bar M_6$-modulated interaction term gives rise to a flat shape for the three point function \cite{b} which contributes to enlarge the allowed classification of bispectra shape-functions for single-field models of inflation (with Bunch-Davies vacuum).\\ 
\subsection{Contact interaction diagram}
We now turn to the calculation of various terms that contribute to the contact interaction diagrams. The terms driven by $M_3, M_4$ are DBI-generated and have been calculated in a number of papers, notably \cite{trispectrum,chen-tris}. $M_2$ is found in both DBI and Ghost inflationary theories \cite{chen-tris, muko}. If we are to preserve both symmetries then we need to focus on $\bar M_{11},\bar M_{12}, \bar N_{1}..\bar N_{5}$ as one can easily check from \textit{Table 1}.\\ Given any interaction term at fourth order, its contribution to the contact interaction diagram can be written as:

\bea
\fl \langle \pi_{k_1}\pi_{k_2}\pi_{k_3}\pi_{k_4}(t\rightarrow 0)\rangle ^{c.i.}_{H_4} = 
-2 \mathcal{I}_m \Big[ \langle 0|\bar T \{ i\int_{-\infty}^{t\rightarrow 0}{ d^3 x \,\, dt^{'} \mathcal{H}_4(x) \pi^{*}_{k1}\pi^{*}_{k2}\pi^{*}_{k3}\pi^{*}_{k4}(t\rightarrow 0)}\}|0\rangle \Big]  \nonumber\\
\eea

\noindent where $\mathcal{I}_m$ stands for taking the imaginary part; a delta enforcing momentum conservation and unimportant numerical factors have been omitted.\\
We have applied the above formula to a number of fourth-order interaction terms providing some examples of notable {\bf S1} \textit{and} {\bf S2}-abiding terms, {\bf S1} \textit{ or} {\bf S2} invariant contributions and, finally, terms that do not respect any of the symmetries above. We start with the {\bf S2}-invariant interaction term $\sim M_2^4 (\nabla \pi)^4$. This contribution is present in both DBI-like and Ghost inflationary models (e.g. in DBI one simply has  $1/c_s^2 = 1-2 M_2^4/M_P^2 \dot H$). Although already written down in \cite{muko, chen-tris}, this term has not been plotted in all four configurations described in section 4 we employ here. This is because in DBI-like theories it is expected to be subdomimant with respect to the $M_4^4 {\dot \pi}^4$ term. \\

$\bullet$  {\bf $ {\cal O}_{1}=1/2\,\,   M_2^4 \, (\partial_{i} \pi)^4 \,\,/ a^4 $}\\

\noindent For the shape function of the operator $(\nabla \pi)^4$ we see that a number of interesting issues arise. First, the plot in the equilateral configuration does not resemble any of those plotted in \cite{chen-tris}. \footnote{See \textit{Appendix B} for a detailed account of the plot of this term first done in \cite{muko} with different variables giving results which are essentially identical to ours despite a simplyfing assumption on our part.} Then, as we mentioned in Fig.~\ref{m62}, in the double squeezed configuration the $k_{12} \rightarrow 0$ limit gives a non-zero finite shape function. This is important because, up to the results in \cite{chen-tris}, this limit was thought as very useful to distinguish the leading contributions coming from interactions at third order in perturbations from the ones at fourth order in fluctuations.

\begin{figure}[hp]
	\includegraphics[scale=0.53]{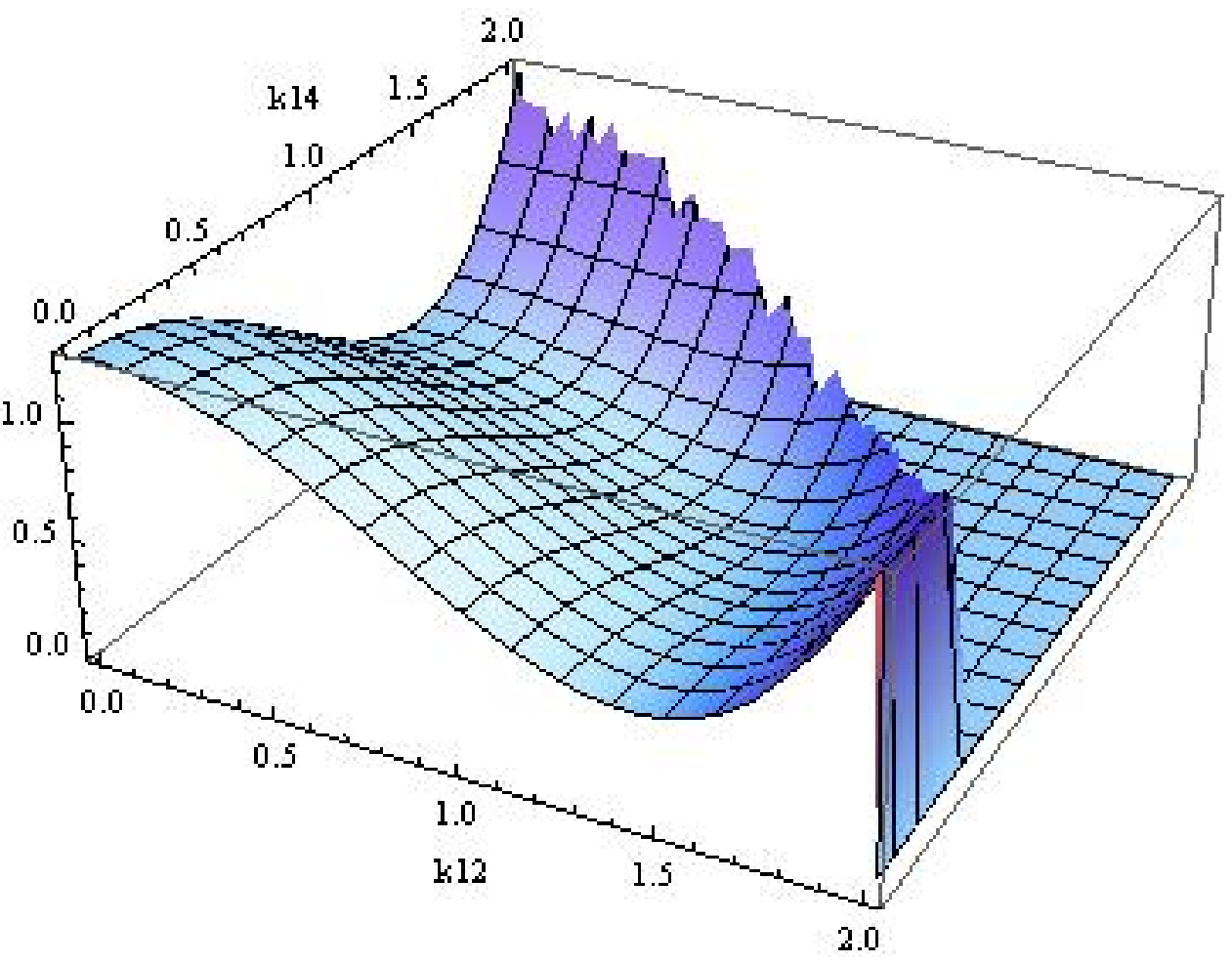}
	\hspace{10mm}
		\includegraphics[scale=0.57]{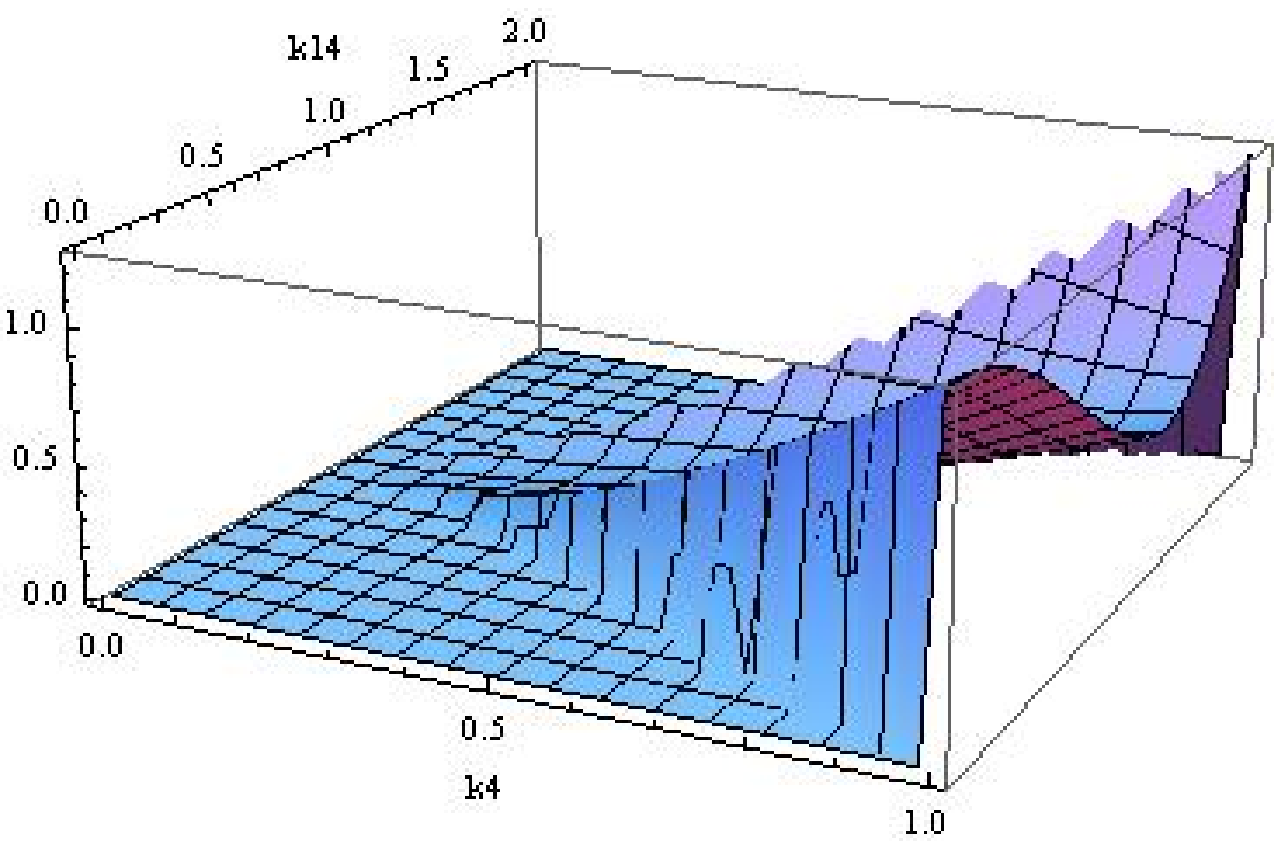}
\caption{The \textit{equilateral} configuration shape for the ${\cal O}_{1}$ operator is presented on the left. It is different from the results in \cite{chen-tris} and very much resembles the plot we obtained for the scalar exchange calculation. Notice that upon performing a change of variables, it is basically identical to the $(\nabla \pi)^4$ interaction term plotted in \cite{muko} (We elaborate further on this point in \textit{Appendix B}).\\ On the right we plotted our findings for the ${\cal O}_{1}$ interaction term in the \textit{folded} configuration. As it will be for the other interactions, this configurations provides no particularly distinctive features that would allow to single out the constributions from the different interaction operators.}
\label{M41}
\end{figure}
\noindent  In fact, all of the terms contributing to the scalar exchange diagram in \cite{chen-tris} give a shape function which in the $k_{12}\rightarrow 0$ is finite. On the contrary, the leading  contact interaction diagram contributions analyzed in \cite{chen-tris} do vanish in this limit. Note also that if one is to relax the assumption of a Bunch-Davies vacuum, the authors of \cite{chen-tris} showed that this is not true anymore. 

\begin{figure}[hp]
	\includegraphics[scale=0.55]{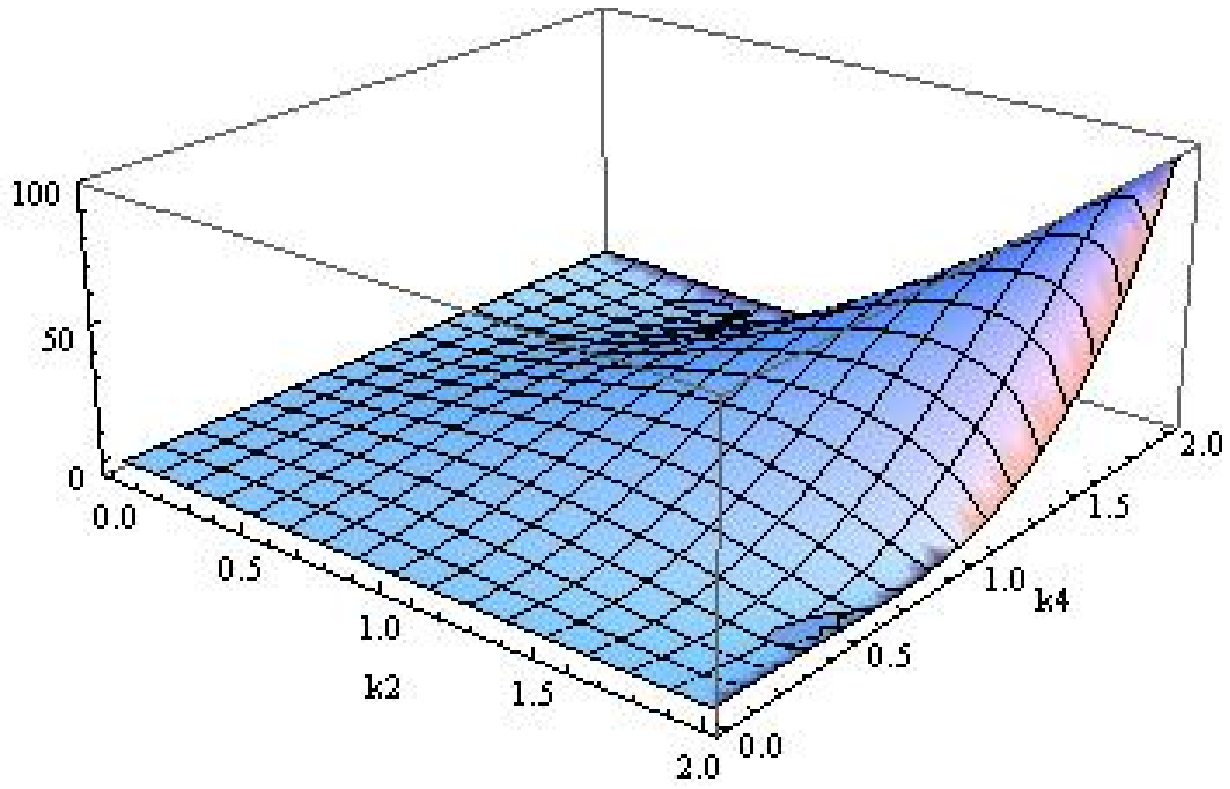}
	\hspace{10mm}
		\includegraphics[scale=0.48]{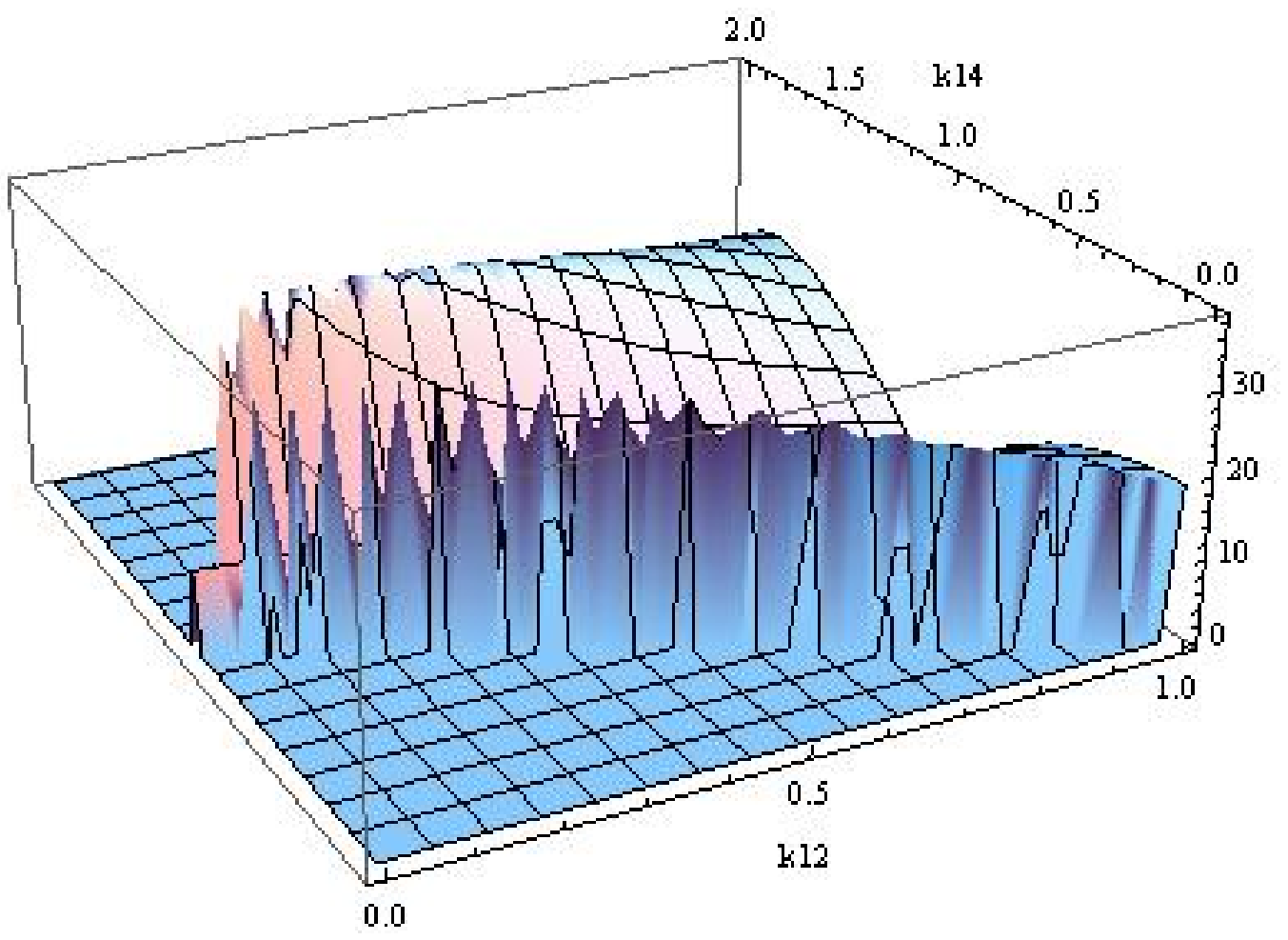}
\caption{The ${\cal O}_{1}$ interaction  \textit{planar} configuration shape  on the left is not exceedingly different from the ones presented in \cite{chen-tris}: it vanishes for $k_2, k_4 \rightarrow 0$, it is peaked for $k_2 = 2 = k_4$. On the $k_2=k_4$ line the shapefunction is convex, rather than concave as for the contact interaction term plotted in \cite{chen-tris}.\\ On the right we plotted the ${\cal O}_{1}$ interaction shape function in the \textit{planar limit double squeezed configuration}.  Here we immediately note an interesting feature: despite this being a contribution to the contact interaction diagram, in the $k_4=k_{12} \rightarrow	0$ it gives a finite, non zero shape function. We comment more on this fact in the text.}
\label{M42}
\end{figure}

\noindent Next, we continue keeping our attention focused on terms which are {\bf S1} \textit{and} {\bf S2} invariant. These include, in terms of their free coefficient, $\bar M_{11}, \bar M_{12}, \bar N_1.. \bar N_5 $. Since they generate shape functions which are qualitatively very similar, we chose to plot just two representative terms in this list. \\

$\bullet$  {\bf $ {\cal O}_{2}=1/6\,\,  \bar M_{11}^2 \, {\dot \pi}^2 \,(\partial_{i}^2 \pi)^2 \,\,/ a^4 $}\\

In the plots below we see that the $\bar M_{11}$-driven term,  generates shapes which are very similar to the ones plotted in \cite{chen-tris} for the DBI-generated term $\sim {\dot \pi}^4$.
\newpage
\begin{figure}[hp]
	\includegraphics[scale=0.53]{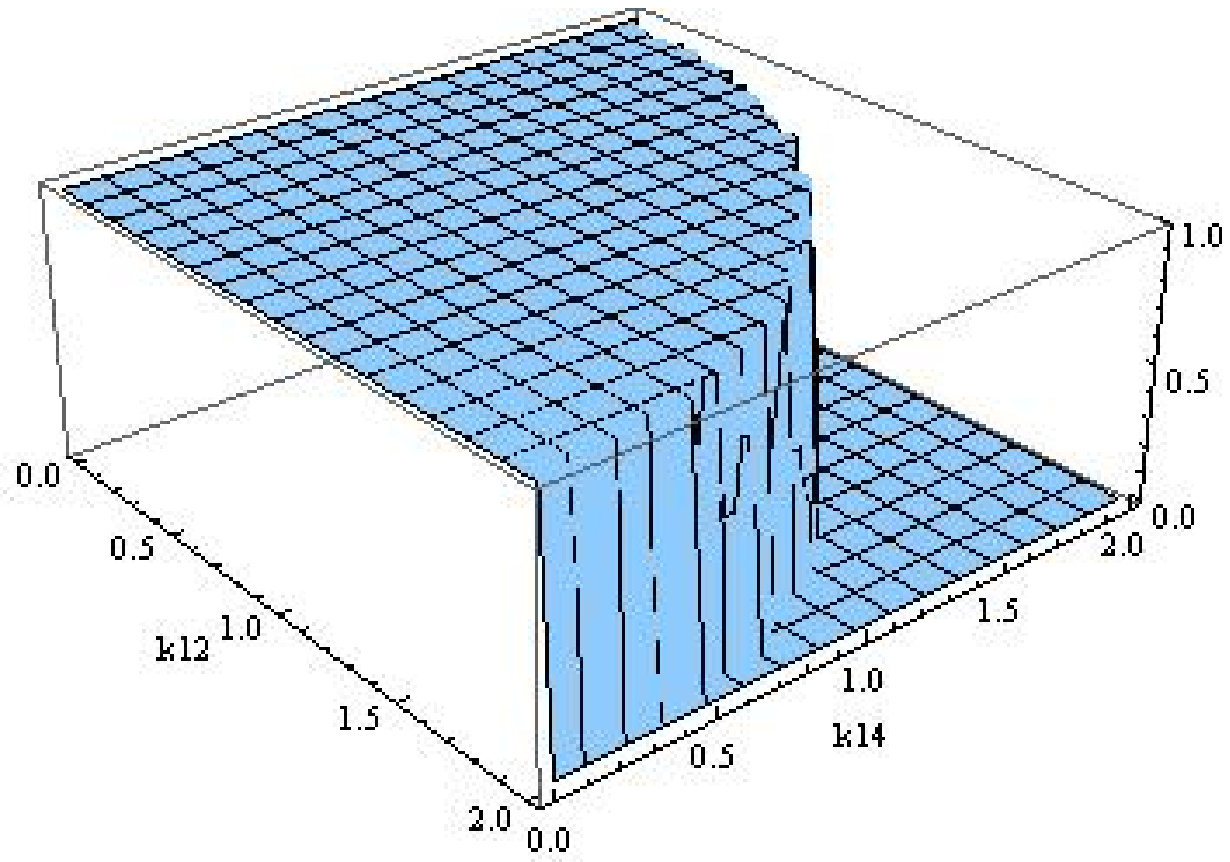}
	\hspace{10mm}
		\includegraphics[scale=0.61]{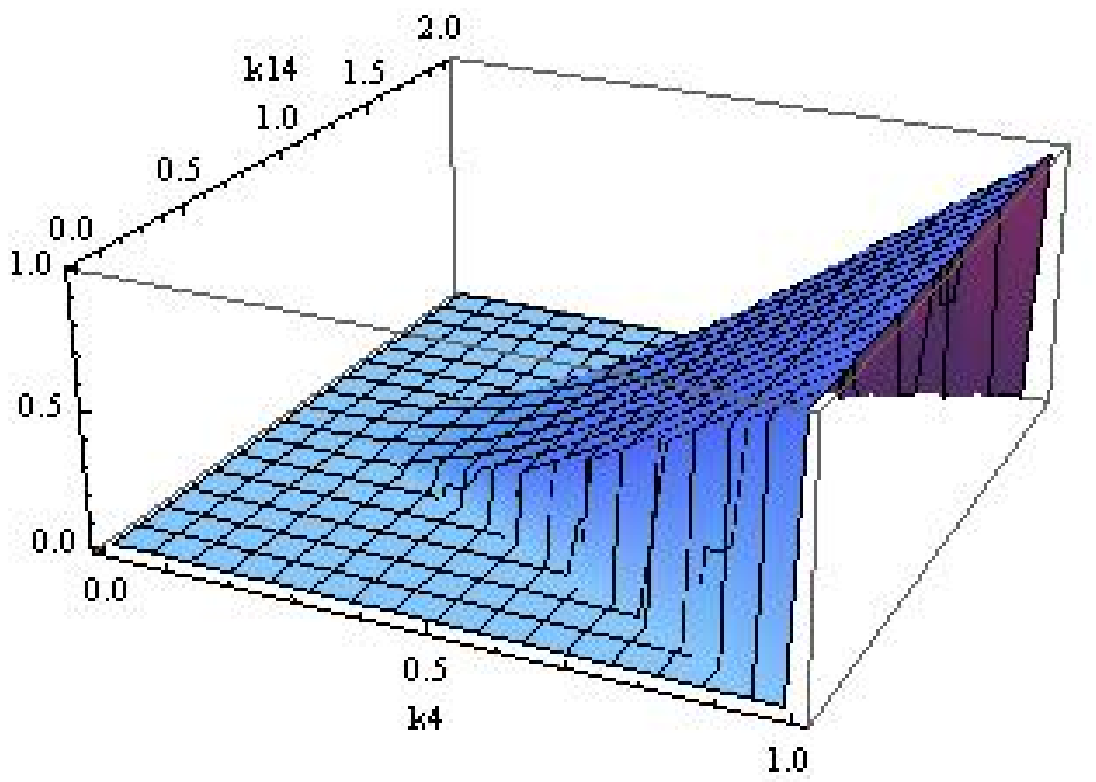}
\caption{The \textit{equilateral} configuration shape is presented on the left for the $ {\cal O}_{2}$ operator. Because of the way the space derivatives are written in Fourier space there's no $k_{12}, k_{14}$ dependence and so one gets a plateau. On the right our findings for $ {\cal O}_{2}$ in the \textit{folded} configuration.}
\label{M111}
\end{figure}

\begin{figure}[hp]
	\includegraphics[scale=0.61]{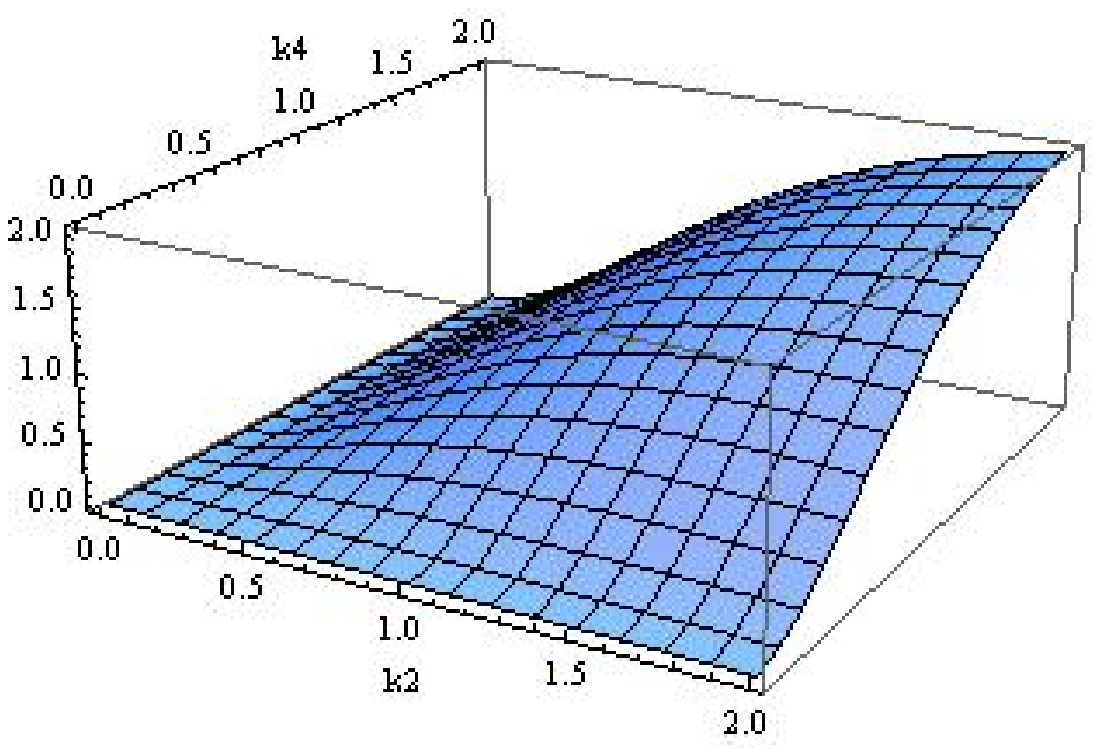}
	\hspace{10mm}
		\includegraphics[scale=0.47]{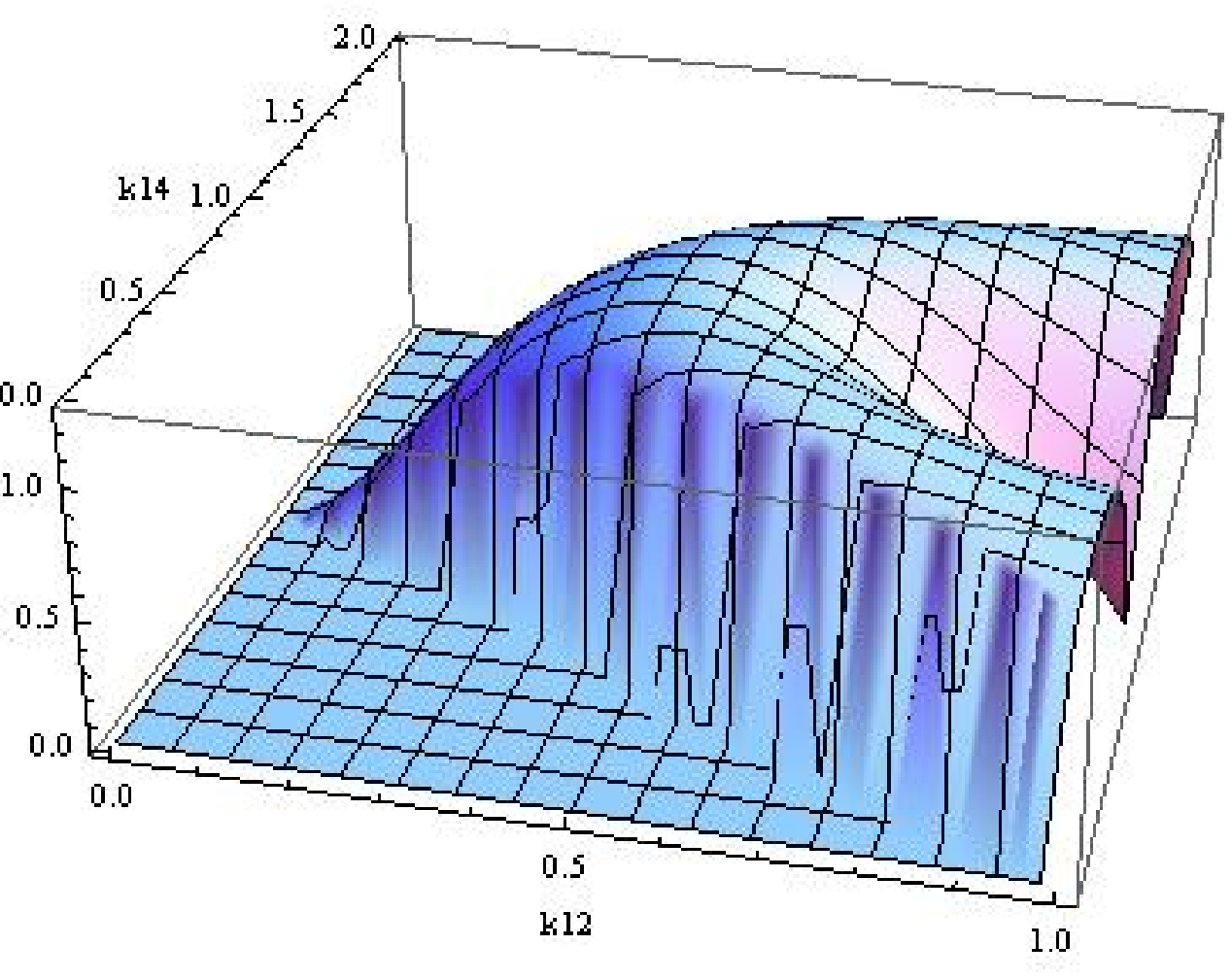}
\caption{The $ {\cal O}_{2}$ \textit{planar} configuration shape  on the left: it vanishes for $k_2, k_4 \rightarrow 0$, it is peaked for $k_2 = 2 = k_4$. On the $k_2=k_4$ line the shapefunction is now concave, just like one would get for the ${\dot \pi}^4$ interaction. \\ On the right we plotted the shape function in the \textit{planar limit double squeezed configuration} associated to the $ {\cal O}_{2}$ interaction term.}
\label{M112}
\end{figure}

\noindent We proceed with the other representative term:\\

$\bullet$  {\bf $ {\cal O}_{3}=1/4!\,\,  \bar N_3 \, \partial_{\rho}^2 \pi \,  \partial_{ij} \pi \, \partial_{jk} \pi \, \partial_{ki} \pi\,\,/ a^8 $}\\

The differences with respect to the $\bar M_{11}$-driven interaction shapes are to be find in the first and the third configuration: in the first configuration they are due to the $k$-dependence of the interaction, on the third configuration $N_3$ gives a plot similar to the one tuned by the $M_2$ coefficient(see Fig.~ \ref{M42}).
\newpage
\begin{figure}[hp]
	\includegraphics[scale=0.62]{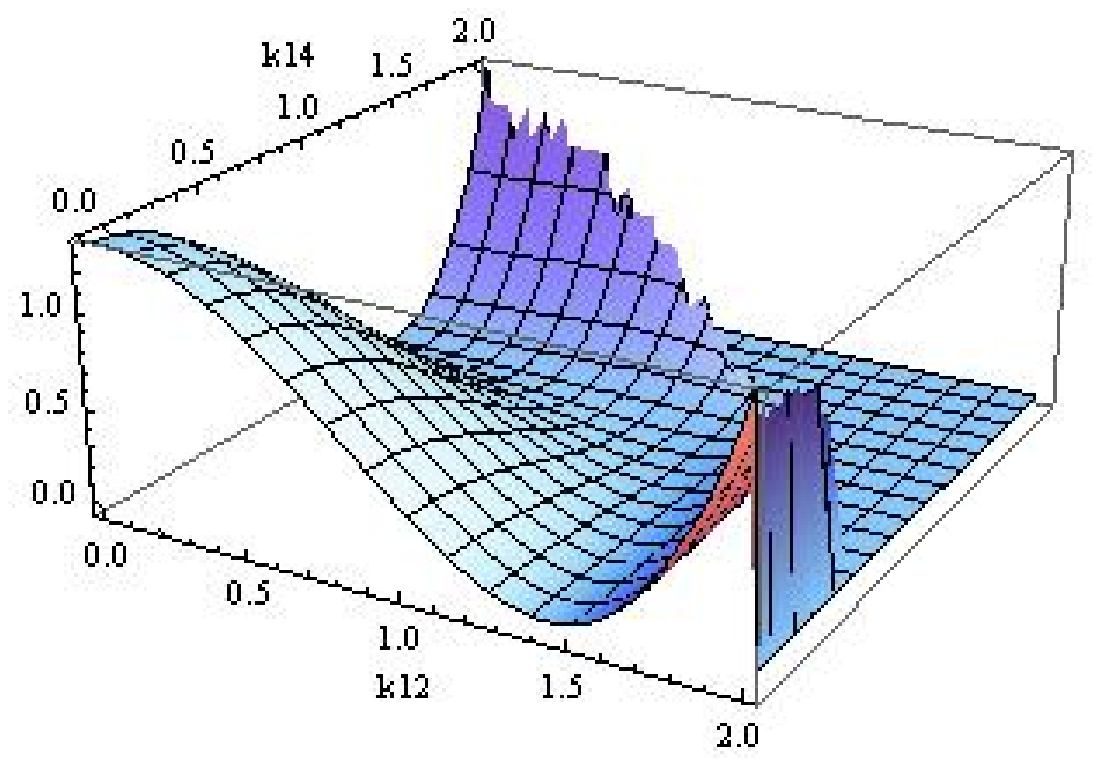}
	\hspace{10mm}
		\includegraphics[scale=0.55]{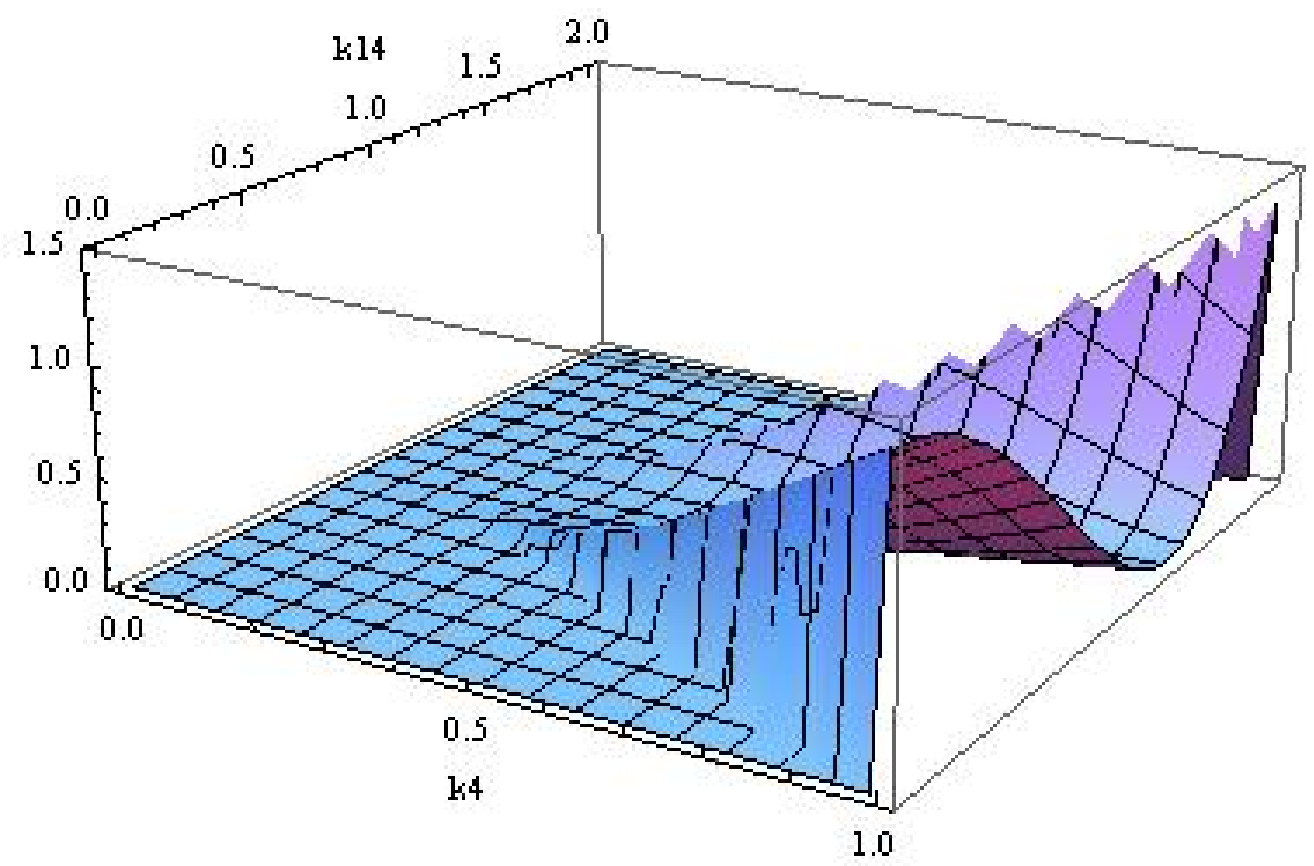}
\caption{The \textit{equilateral} configuration shape for $ {\cal O}_{3}$ is presented on the left. On the right our findings for the $ {\cal O}_{3}$ operator in the \textit{folded} configuration.}
\label{N31}
\end{figure}

\begin{figure}[hp]
	\includegraphics[scale=0.59]{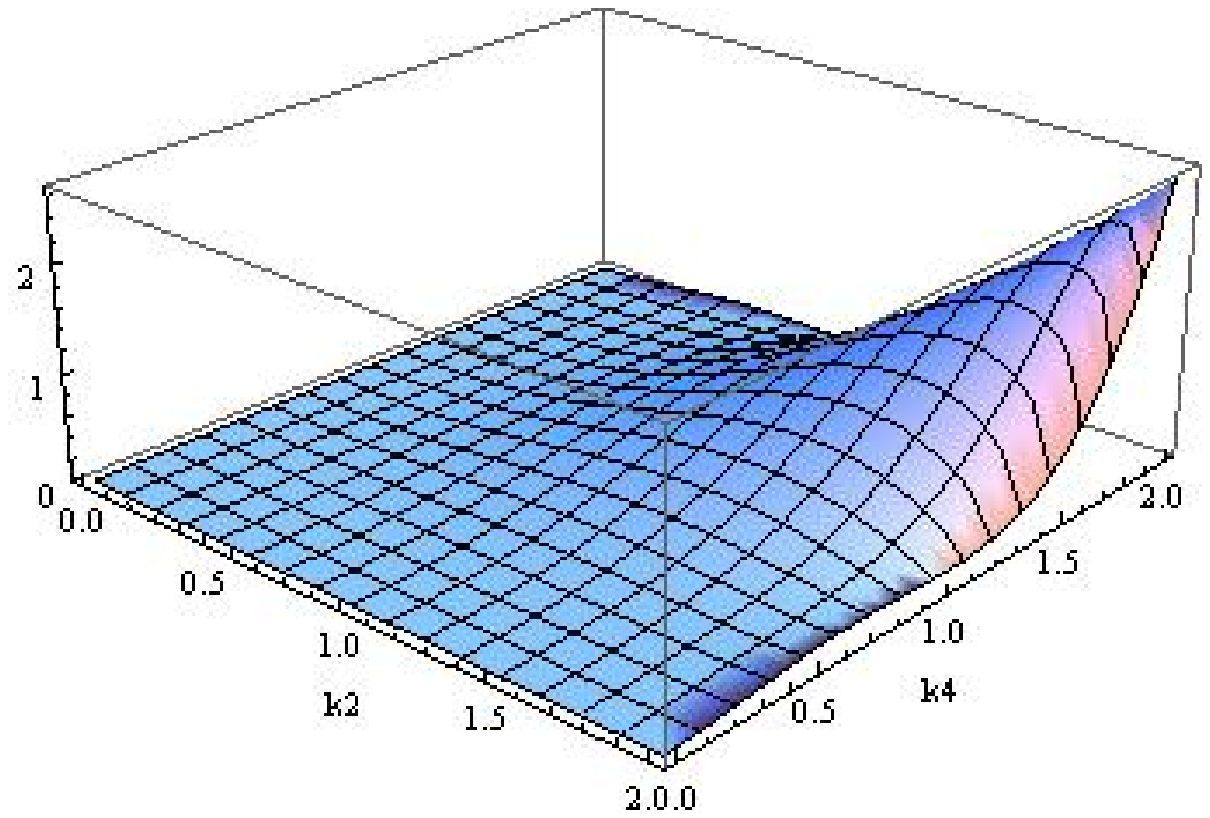}
	\hspace{10mm}
		\includegraphics[scale=0.57]{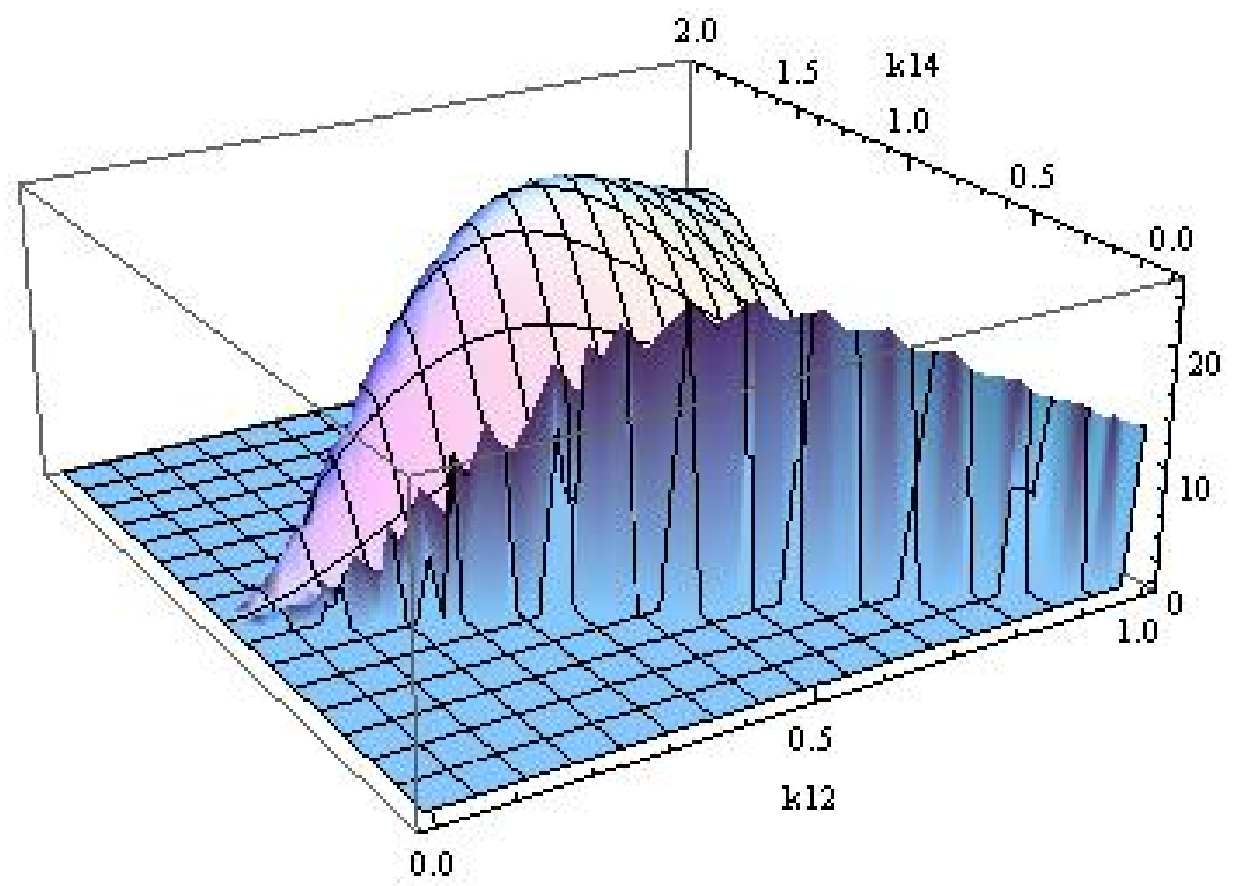}
\caption{The \textit{planar} configuration shape  on the left for $ {\cal O}_{3}$. On the right we plotted the shape function of the $ {\cal O}_{3}$ interaction term in the \textit{planar limit double squeezed configuration}.}
\label{N32}
\end{figure}

\noindent We now turn our attention onto terms which violate one of the symmetries, {\bf S1} in this case. Indeed, we analyze the interaction $ \sim M_2^4 M_3^4 {\dot \pi}^2 (\partial_{i} \pi)^2\,/a^2$\\

$\bullet$  {\bf $ {\cal O}_{4}= (M_2^4 M_3^4)/(2M_2^4 +M_{P}^2 \epsilon H^2 -3\bar M_1^3 H)\,\times  {\dot \pi}^2 \, ( \partial_{i} \pi)^2\, / a^2 $}\\
\newpage
\begin{figure}[hp]
	\includegraphics[scale=0.60]{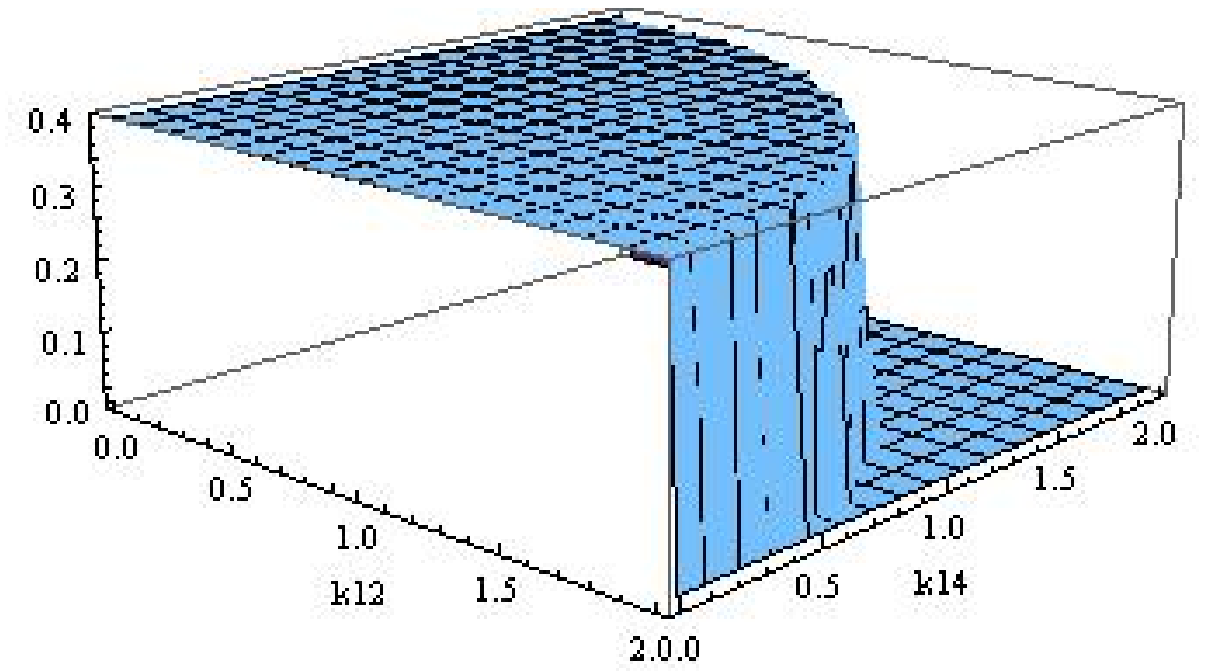}
	\hspace{10mm}
		\includegraphics[scale=0.56]{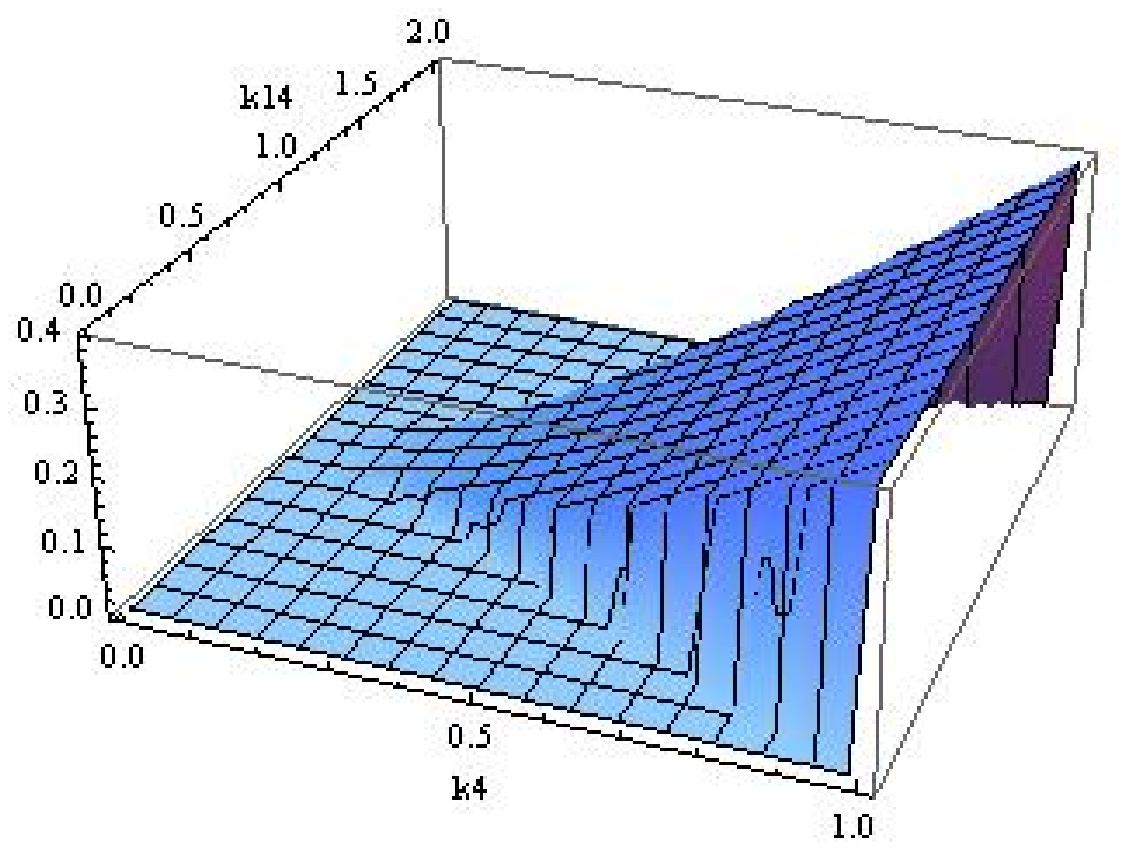}
\caption{The \textit{equilateral} configuration shape for $ {\cal O}_{4}$ is presented on the left. On the right our findings for the \textit{folded} configuration associated to $ {\cal O}_{4}$. Both are very similar to we obtained for the $\bar M_{11}$ in Fig.~\ref{M111} and to what was found for the $M_4^4$-driven interaction in the literature \cite{chen-tris}.}
\label{M231}
\end{figure}

For this interaction term we see the interesting feature presents itself in the fourth configuration where the $k_{12} \rightarrow 0$ limit gives a finite shape function.

\begin{figure}[hp]
	\includegraphics[scale=0.60]{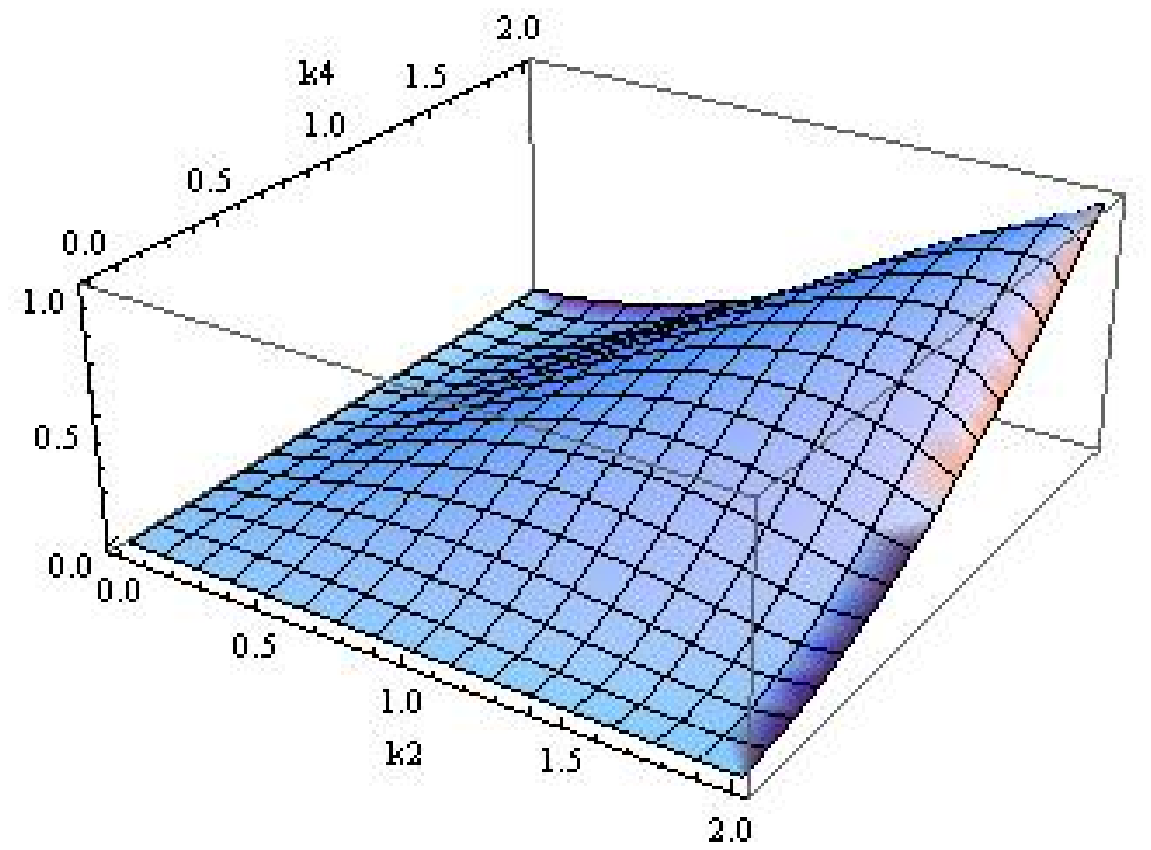}
	\hspace{10mm}
		\includegraphics[scale=0.52]{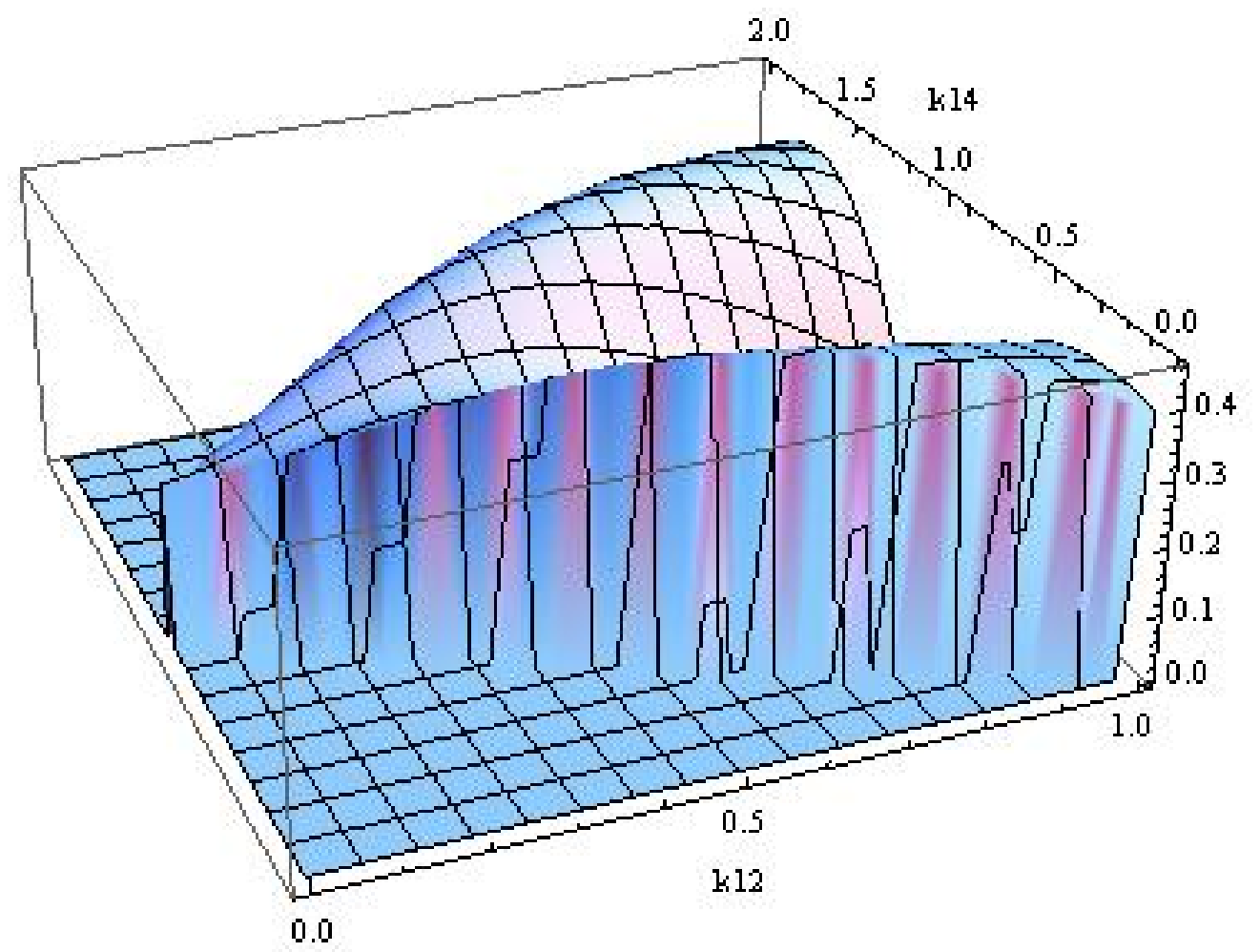}
\caption{The \textit{planar} configuration shape for $ {\cal O}_{4}$  on the left. On the right we plotted the $ {\cal O}_{4}$-generated shape function in the \textit{planar limit double squeezed configuration} which gives again a non-zero and finite shape function even for fourth-order interactions such as the one under scrutiny here.}
\label{M232}
\end{figure}

We now proceed to plot our findings for one more term, precisely the leading fourth-order interaction term among the ones driven by $\bar M_6$. It is clear that whenever this term gives a leading third order contribution (something one can achieve given the freedom on most ${\bf M}_n$'s ), it violates {\bf S1}. {\bf S2} however, is preserved by the leading terms associated to this coefficient at third and fourth order as is clear from Eq.~(\ref{h4}) and \textit{Table 1}.\\

$\bullet$  {\bf $ {\cal O}_{5}=1/6\,\, \bar M_6   (\partial_k \pi)^2 \, ( \partial_{ij} \pi)^2\, / a^6 $}\\
\newpage
\begin{figure}[hp]
	\includegraphics[scale=0.54]{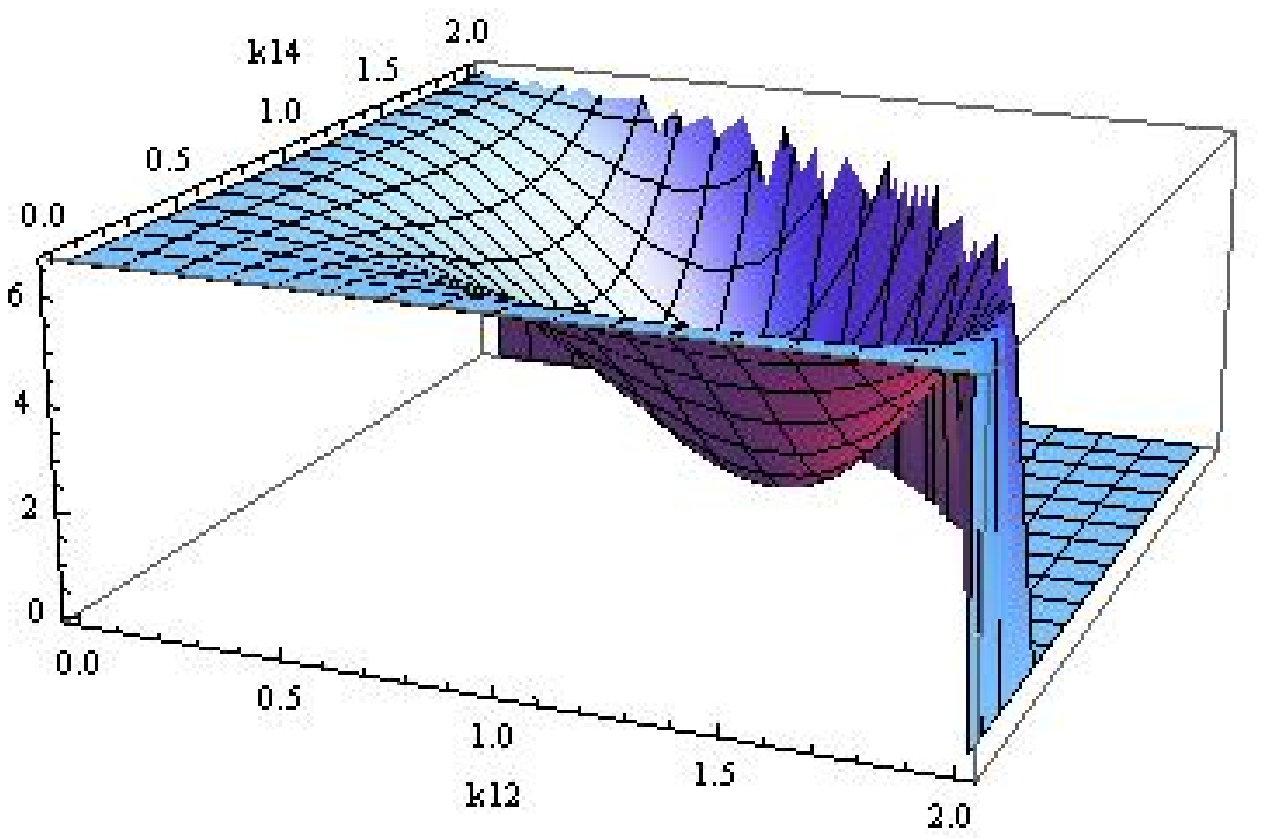}
	\hspace{10mm}
		\includegraphics[scale=0.60]{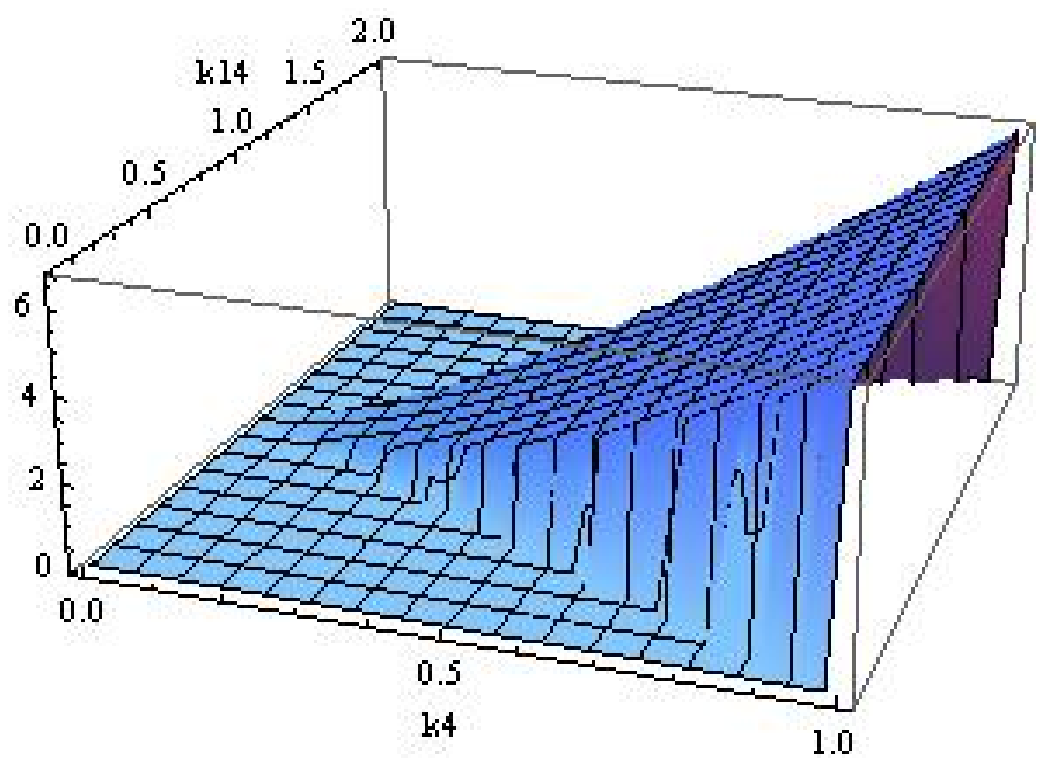}
\caption{The \textit{equilateral} configuration shape for the operator $ {\cal O}_{5}$ is presented on the left: this shape for the plot function has not been seen before in the equilateral configuration.\\ On the right our findings for the $ {\cal O}_{5}$ interaction term in the \textit{folded} configuration.}
\label{M641}
\end{figure}

\begin{figure}[hp]
	\includegraphics[scale=0.64]{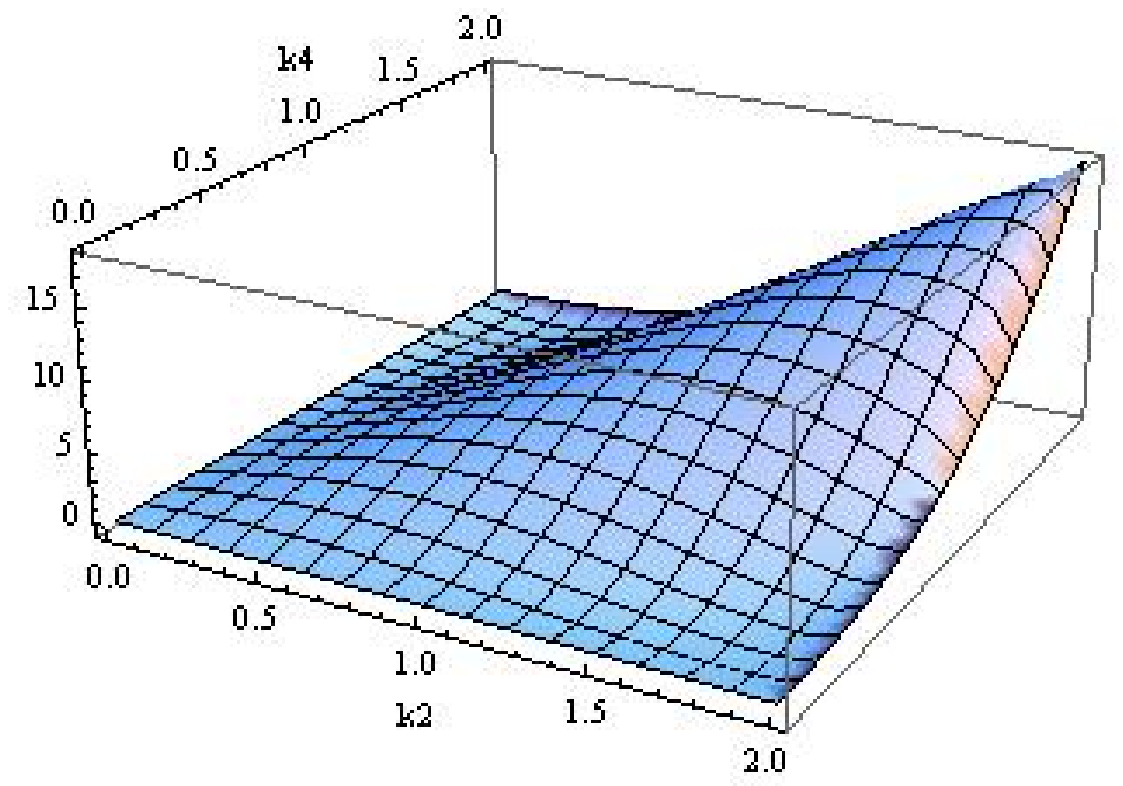}
	\hspace{10mm}
		\includegraphics[scale=0.51]{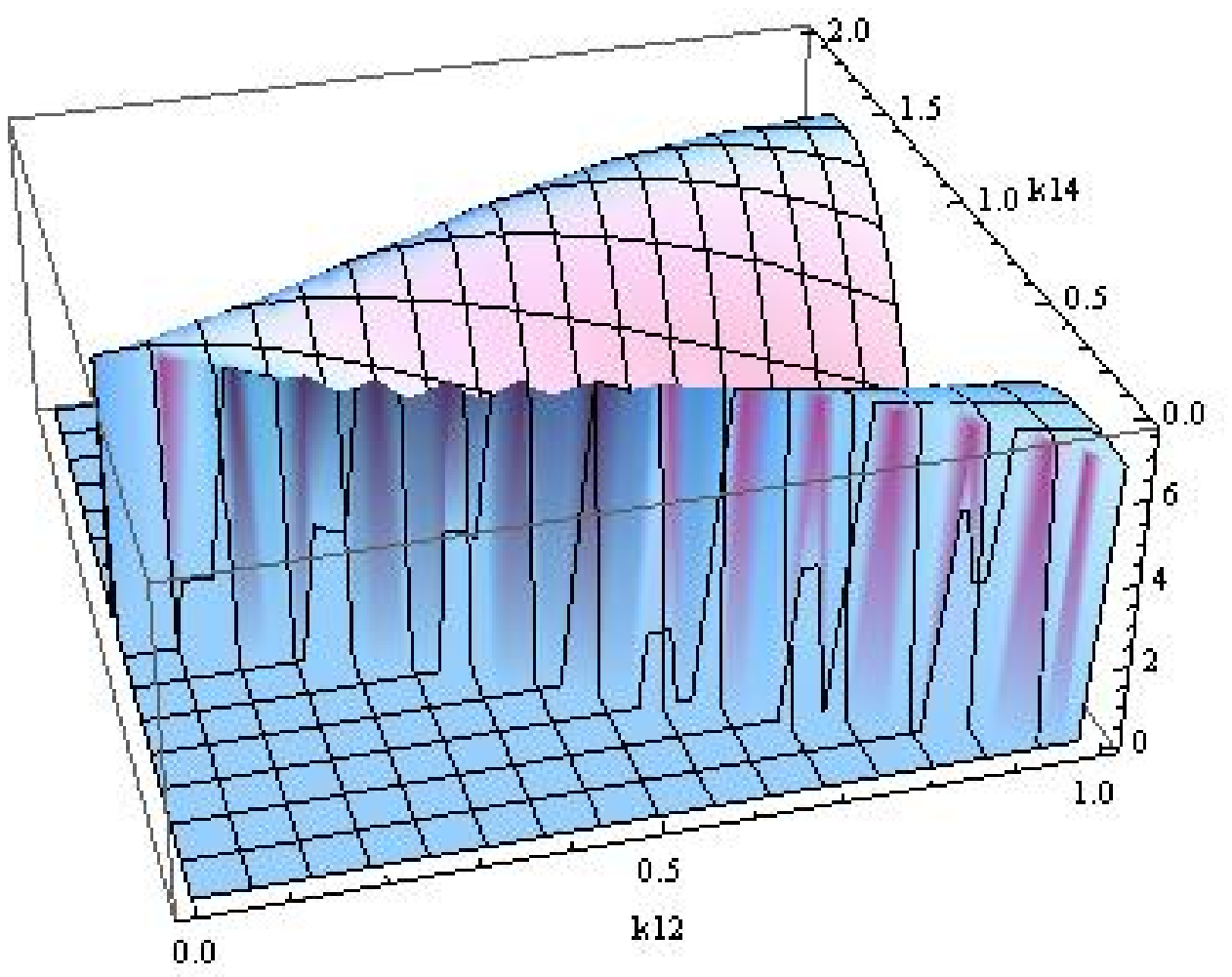}
\caption{The \textit{planar} configuration shape for $ {\cal O}_{5}$  on the left. On the right we plotted the $ {\cal O}_{5}$-generated shape function in the \textit{planar limit double squeezed configuration} }
\label{M642}
\end{figure}

As one can see from Fig.~\ref{M641}, the plot in the equilateral configuration has no analogue in the shapes of \cite{chen-tris, muko} for this configuration. It is somewhat reminiscent of the shape obtained for the ${\dot \pi}^4$ of \cite{chen-tris} but again, we stress it was obtained in a different configuration. The results plotted in Fig.~\ref{M642} show once again that it is not safe in theories more general than DBI to attribuite to the \textit{planar limit double squeezed} configuration the role to provide a distinctive signature in the $k_{12} \rightarrow 0$ limit that would enable one to distinguish between third and fourth-order interaction contributions (see the discussion in \cite{chen-tris}).

\newpage

\section{Conclusions}

In this work we aimed at employing the tools of effective field theory \cite{eft08} to probe the space of all possible interactions for inflation driven by a single scalar degree of freedom up to fourth order in perturbations. Following a companion work on the bispectrum \cite{b}, and guided by the requirement of some additional symmetries on the action as an ordering principle \cite{muko,4pt}, in \textit{Section 4} we singled out the distinctive features in the trispectrum one obtains when considering curvature-generated terms from a very general fourth-order Hamiltonian. It is important to note that all of these intereactions allow, by construction, for a large trispectrum. Some of them present features which also emerge in DBI-inflation and Ghost inflation \cite{chen-tris, muko}.\\ We have also identified third and fourth-order interaction terms which have not been analyzed before and whose analysis reveals novel interesting effects. We were able to show  that, unlike what happens in DBI-like models,
the analysis of the \textit{double squeezed} configuration cannot give a clear cut clue as to what kind of signal comes from leading third-order terms as opposed to fourth-order terms in perturbations. We found many interactions generating a shape in the equilateral configuration that mimics the behaviour of the ghost interaction term ( i.e. $(\nabla \pi)^4$ ) shape function first plotted in \cite{muko} which is quite different from the shapes of the DBI model (we also extended the ghost inflation plots of \cite{muko} to three other configurations). Triggered by an interesting flat shape found in \cite{b}, specifically the one controlled by the $\bar M_6$ coefficient, the analysis of this term contributions to the scalar exchange and contact interaction diagram was calculated and plotted: a shape function which has not been found before emerged in the equilateral configuration for the contact interaction contribution. Finally, in 
\textit{Section 4}, we again verified up to fourth order a feature which already emerged at the level of the bispectrum in \cite{b}: a shape function which in \textit{general single field inflation} models  is only obtained either by employing a linear combination of operators (as far as the Bispectrum is concerned) or relaxing the Bunch-Davies vacuum requirement for the theory, quite naturally arises in more general setups as the one employed here. Furthermore, it does so when considering several and independent interaction terms. 

\section*{Acknowledgments}
It is a pleasure to thank Xingang Chen and Leonardo Senatore for sharing very useful comments on the manuscript.
This research has been partially supported by the ASI Contract No. I/016/07/0 COFIS, the ASI/INAF Agreement I/072/09/0 for the Planck LFI Activity of Phase E2. MF would like to thank Claudio Destri for fruitful discussions and for kind encouragement during the completion of this work. MF is happy to thank the Physics Department of the University of Padova for warm hospitality.
\section{Appendix A}
 A more detailed presentation of the scalar exchange calculation for the $\bar M_6$-driven term is presented here. We start from the scalar exchange part of  Eq.~(\ref{z4}):
 \bea
 \langle \Omega|\zeta_{k_1}\zeta_{k_2}\zeta_{k_3}\zeta_{k_4}(t)|\Omega\rangle_{s.e.}= \nonumber\\ \langle 0|\bar T \{i\int_{-\infty}^{t_0} d^3 x dt^{'} \mathcal{H}_3(x)\} \zeta_{k_1}\zeta_{k_2}\zeta_{k_3}\zeta_{k_4}(t)\,  T\{ -i\int_{-\infty}^{t_0} d^3 x^{'} dt^{''} \mathcal{H}_3(x) \}|0\rangle \nonumber\\
 +\langle 0|\bar T \{\frac{i^2}{2}\int{ \int{ d^3 x \,dt\, d^3 x^{'}\, dt^{''} \mathcal{H}_3(x) \mathcal{H}_3(x^{'})}}\} \zeta_{k_1}\zeta_{k_2}\zeta_{k_3}\zeta_{k_4}(t)|0\rangle \nonumber\\
 +\langle 0|\zeta_{k_1} \zeta_{k_2}\zeta_{k_3}\zeta_{k_4}(t) T \{\frac{(-i)^2}{2}\int_{-\infty}^{t_0}{ \int_{-\infty}^{t_0}{ d^3 x \,dt\, d^3 x^{'}\, dt^{''} \mathcal{H}_3(x) \mathcal{H}_3(x^{'})}}\}|0\rangle .
\nonumber\\ \label{sed}
\eea
Using Wick contraction on a generic operator  $\phi$, one has:
\bea
\fl T\{\phi(t_1)\phi(t_2)\phi(t_3)\phi(t_4)\}=N\{\phi(t_1)\phi(t_2)\phi(t_3)\phi(t_4)+\textit{all \,\, contractions}  \},
\eea
 where \textit{N} is the normal ordering operator.
Note also that, being our $\pi \sim \zeta$ operators squeezed between two vacua of the free theory, this reduces to considering only terms which are writeable as fully contracted contributions. For the anti-time order operator the same formula holds, only one needs to define contractions differently. We show below this difference:
\bea
\fl \contraction{\phi(}{\vec x_1, t_1) \phi( }{ \vec x_2, t_2},
\phi(\vec x_1 , t_1)\,\,\,\, \phi(\vec x_2, t_2)_{\bf{T}} =[\phi^{+}(\vec x_1 ,t_1), \phi^{-}(\vec x_2,t_2)] \theta(t_1-t_2)+[\phi^{+}(\vec x_2 ,t_2), \phi^{-}(\vec x_1,t_1)] \theta(t_2-t_1) \nonumber\\
\eea

\bea
\fl \contraction{\phi(}{\vec x_1, t_1) \phi( }{ \vec x_2, t_2},
\phi(\vec x_1 , t_1)\,\,\,\, \phi(\vec x_2, t_2)_{{\bf \bar T}} =[\phi^{+}(\vec x_2 ,t_2), \phi^{-}(\vec x_1,t_1)] \theta(t_1-t_2)+[\phi^{+}(\vec x_1 ,t_1), \phi^{-}(\vec x_2,t_2)] \theta(t_2-t_1) \nonumber\\
\eea

where 
\be
\fl \phi^{+}(\vec x_2 ,t_2)= \int{\frac{d^3 k}{(2\pi)^3} \phi(\vec k,t_2)\, {\bf a_k}e^{i \vec{k} \cdot \vec x_2}}; \quad  \phi^{-}(\vec x_2 ,t_2)= \int{\frac{d^3 k}{(2\pi)^3} \phi^{*}(\vec{k},t_2)\, {\bf a^{\dagger}_{k}}e^{-i \vec{k} \cdot \vec x_2}}.
\ee

\noindent Using the definitions above one gets several different contributions from Eq.~(\ref{sed}). Note also that, using time and anti-time order definitions, the last two lines of Eq.~(\ref{sed})  are just each other's conjugate and can therefore be grouped together. We now procede to write an explicit expression for the four point function generated by the $\bar M_6$-driven contribution to the scalar exchange diagram:
\bea
\fl \langle \pi_{k1} \pi_{k2}\pi_{k3}\pi_{k4}\rangle^{s.e.}_{\bar M_6}\propto
\eea
\bea
\fl  \frac{M_2^8}{3}\left[ 4\cdot \left(\pi^{*}_{k1}\pi^{*}_{k2}\pi_{k3}\pi_{k4}(0)\,\int_{-\infty}^{0}{dt_1  \frac{a^3}{a^4} \dot \pi_{k12} \pi_{k1}\pi_{k2}({\bf k}_1\cdot {\bf k}_2)^2 }\int_{-\infty}^{0}{dt_2 \frac{a^3}{a^4} \dot \pi^{*}_{k12} \pi^{*}_{k3}\pi^{*}_{k4}({\bf k}_3\cdot {\bf k}_4)^2} \right. \right. \nonumber \\
\fl \left. \left. +\textit{ 5}\,\,\textit{ permutations} \right)+\right. \nonumber \\
\fl \left. \left( 2 \cdot 2 \cdot\,  \pi^{*}_{k1}\pi^{*}_{k2}\pi_{k3}\pi_{k4}(0)\,\int_{-\infty}^{0}{dt_1  \frac{a^3}{a^4} \dot \pi_{k1} \pi_{k12}\pi_{k2}(-{\bf k}_{12}\cdot {\bf k}_2)^2} \int_{-\infty}^{0}{dt_2 \frac{a^3}{a^4} \dot \pi^{*}_{k12} \pi^{*}_{k3}\pi^{*}_{k4} ({\bf k}_3\cdot {\bf k}_4)^2 } \right. \right. \nonumber \\
\left. \left. \fl +\textit{11}\,\,\textit{ permutations}  \right) \right.\nonumber \\
\left.\fl    \left( 2 \cdot 2 \cdot\,  \pi^{*}_{k1}\pi^{*}_{k2}\pi_{k3}\pi_{k4}(0)\,\int_{-\infty}^{0}{dt_1  \frac{a^3}{a^4} \dot \pi_{k12} \pi_{k1}\pi_{k2}({\bf k}_{1}\cdot {\bf k}_2)^2} \int_{-\infty}^{0}{dt_2 \frac{a^3}{a^4} \dot \pi^{*}_{k3} \pi^{*}_{k12}\pi^{*}_{k4} ({\bf k}_{12}\cdot {\bf k}_4)^2 } \right.     \right.\nonumber\\
\left. \left. \fl  +\textit{11}\,\,\textit{ permutations}  \right)    \right. \nonumber \\
\left. \fl   \left( 4 \cdot   \pi^{*}_{k1}\pi^{*}_{k2}\pi_{k3}\pi_{k4}(0)\,\int_{-\infty}^{0}{dt_1  \frac{a^3}{a^4} \dot \pi_{k1} \pi_{k12}\pi_{k2}(-{\bf k}_{12}\cdot {\bf k}_2)^2} \int_{-\infty}^{0}{dt_2 \frac{a^3}{a^4} \dot \pi^{*}_{k3} \pi^{*}_{k12}\pi^{*}_{k4} ({\bf k}_{12}\cdot {\bf k}_4)^2 } \right.    \right. \nonumber \\
\left. \left. \fl    +\textit{23}\,\,\textit{ permutations}  \right)  \right]+ \nonumber
\eea
\bea
\fl  -\frac{2\,M_2^8}{3} \mathcal{R}_e  \left[ 4\cdot \left(\pi^{*}_{k1}\pi^{*}_{k2}\pi^{*}_{k3}\pi^{*}_{k4}(0)\,\int_{-\infty}^{0}{dt_1  \frac{a^3}{a^4} \dot \pi^{*}_{k12} \pi_{k1}\pi_{k2}({\bf k}_1\cdot {\bf k}_2)^2 }\int_{-\infty}^{t_1}{dt_2 \frac{a^3}{a^4} \dot \pi^{}_{k12} \pi^{}_{k3}\pi^{}_{k4}({\bf k}_3\cdot {\bf k}_4)^2} \right. \right. \nonumber \\
\fl \left. \left. +\textit{ 5}\,\,\textit{ permutations} \right)+\right. \nonumber \\
\fl \left. \left( 2 \cdot 2 \cdot\,  \pi^{*}_{k1}\pi^{*}_{k2}\pi^{*}_{k3}\pi^{*}_{k4}(0)\,\int_{-\infty}^{0}{dt_1  \frac{a^3}{a^4} \dot \pi_{k1} \pi^{*}_{k12}\pi_{k2}(-{\bf k}_{12}\cdot {\bf k}_2)^2} \int_{-\infty}^{t_1}{dt_2 \frac{a^3}{a^4} \dot \pi^{}_{k12} \pi^{}_{k3}\pi^{}_{k4} ({\bf k}_3\cdot {\bf k}_4)^2 } \right. \right. \nonumber \\
\left. \left. \fl +\textit{11}\,\,\textit{ permutations}  \right) \right.\nonumber \\
\left.\fl    \left( 2 \cdot 2 \cdot\,  \pi^{*}_{k1}\pi^{*}_{k2}\pi^{*}_{k3}\pi^{*}_{k4}(0)\,\int_{-\infty}^{0}{dt_1  \frac{a^3}{a^4} \dot \pi^{*}_{k12} \pi_{k1}\pi_{k2}({\bf k}_{1}\cdot {\bf k}_2)^2} \int_{-\infty}^{t_1}{dt_2 \frac{a^3}{a^4} \dot \pi^{}_{k3} \pi^{}_{k12}\pi^{}_{k4} ({\bf k}_{12}\cdot {\bf k}_4)^2 } \right.     \right.\nonumber\\
\left. \left. \fl  +\textit{11}\,\,\textit{ permutations}  \right)    \right. \nonumber \\
\left. \fl   \left( 4 \cdot   \pi^{*}_{k1}\pi^{*}_{k2}\pi^{*}_{k3}\pi^{*}_{k4}(0)\,\int_{-\infty}^{0}{dt_1  \frac{a^3}{a^4} \dot \pi_{k1} \pi^{*}_{k12}\pi_{k2}(-{\bf k}_{12}\cdot {\bf k}_2)^2} \int_{-\infty}^{t_1}{dt_2 \frac{a^3}{a^4} \dot \pi^{}_{k3} \pi^{}_{k12}\pi^{}_{k4} ({\bf k}_{12}\cdot {\bf k}_4)^2 } \right.    \right. \nonumber \\
\left. \left. \fl    +\textit{23}\,\,\textit{ permutations}  \right)  \right]. \label{appb}
\eea
One then performs these calculations and plots the results to obtain Fig.~\ref{m61},\ref{m62}. The situation for the contact interaction diagram contributions is considerably simpler as there is just one time intergral to be performed and two less fields to be taken into account.

\section{Appendix B}
We want here to show with an example what seems to be a general feature concerning the use of (reasonably) approximated wavefunctions in the calculation of higher order correlators. In \cite{b} we found that in  performing an exact calculation for correlators in a very general theory such as the one we employed in this paper, whenever  a given interaction term was producing a shape function for the trispectrum which one could qualitatively classify as, say, equilateral, so was the calculation  performed with a simplified wavefunction. This is due to two independent reasons. First, we start from the realization that, precisely in the horizon-crossing region, which is where one expects the main contribution to any n-point to come from, the exact general wavefunction \cite{b} and the usual one, $H^{(1)}_{3/2}(\tilde{c_s}k\tau)$, which in these theories is an approximated solution, behave very similarly. Secondly, in \cite{b} we concluded that most of the distinctive effects of the bispectrum where due not to the particular k-modes dependence of the result of the integrals like the one in Eq.~(\ref{calc}),  but on the fraction of that k-dependence that could be taken outside the integral, so on the part of the k-dependence not directly attached to the time behaviour of the wavefunction and which is common to the exact and approximated wavefunction.\\
We now compare the trispectrum shapefunction of a ghost inflation interaction term, $(\nabla \pi)^4$, performed with the exact ghost solution in \cite{muko} with the results we obtain employing the approximated DBI wavefunction, just what we used in obtaining all the shape functions presented here.\\
The interaction reads:
\be
\frac{M_2^4}{2} \sum_{i=1}^{3} \sum_{j=1}^{3} \frac{(\partial_i \pi)^2(\partial_j \pi)^2}{a^4}
\ee
Its trispectrum  shapefunction obtained through the approximated methods has been ploted in Fig. \ref{M41}. In order to compare it with the exact calculation of \cite{muko} we need to change variables and turn to:
\bea
\fl C_2= \hat{k}_1 \cdot \hat{k}_2;\,\,\, C_3= \hat{k}_1 \cdot \hat{k}_3;\,\,\, C_4=-1-C_2-C_3; \quad k_{12}= \sqrt{2(1+C_2)}\quad k_{14}=  \sqrt{2-C_2 -C_3)} \nonumber \\
\eea
We now show the plots obtained by performing this change of variable on our approximated result alongside the plot obtained with the exact ghost wavefunction taken directly from \cite{muko}.
\begin{figure}[hp]
	\includegraphics[scale=0.58]{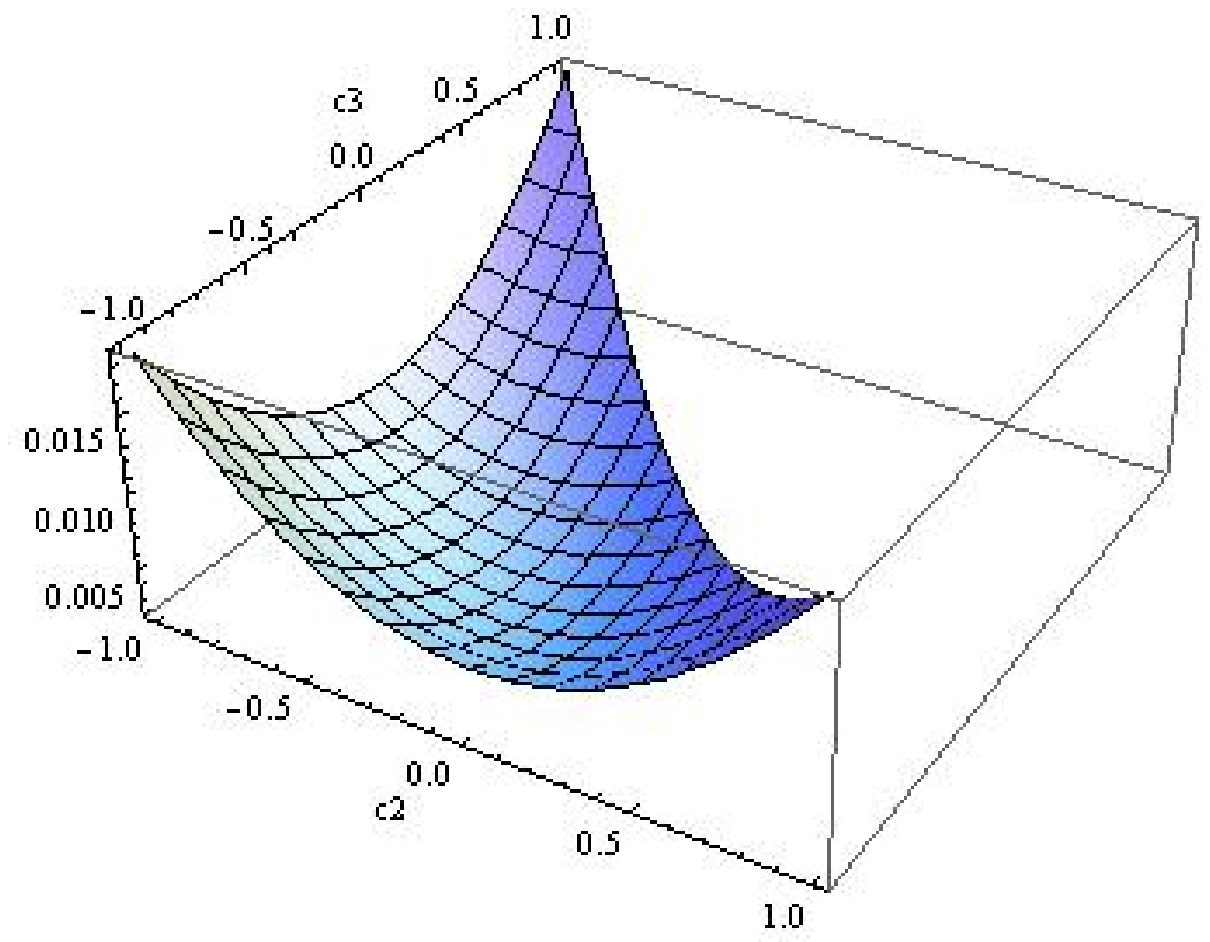}
	\hspace{10mm}
		\includegraphics[scale=0.72]{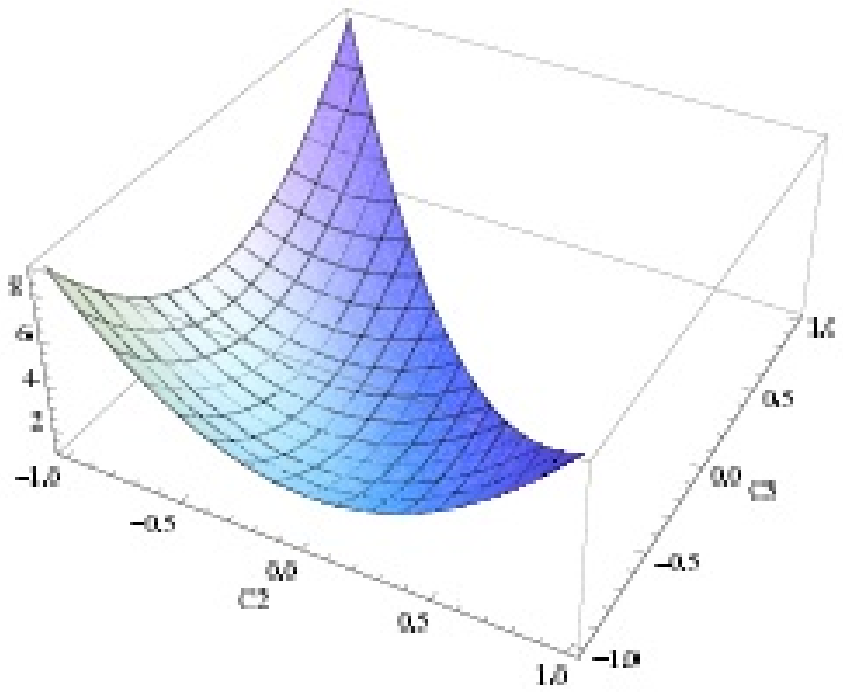}
\caption{On the left the approximated result. The two shapefunctions are qualitatively identical barring an unimportant numerical coefficient due to a different normalization.}
\label{Mghost}
\end{figure}
\newpage
\section*{References}
\bibliographystyle{JHEP}

\end{document}